\documentclass[sigconf,nonacm]{acmart}
\usepackage{defaults}

\usepackage{microtype}
\usepackage[htt]{hyphenat}

\usepackage[normalem]{ulem}
\usepackage{cleveref}
\usepackage{graphicx}
\usepackage[utf8]{inputenc}
\usepackage[noend]{algpseudocode}
\usepackage{listings}
\usepackage{tikz}
\usepackage{makecell}
\usepackage{pifont}

\usepackage{soul}
\soulregister\cite7
\soulregister\ref7
\soulregister\pageref7

\def\name{Remote-Timer-as-a-Service}

\newcommand{\ignore}[1]{}

\usepackage{caption,newfloat}
\usepackage{listings}
\usepackage{color}
\definecolor{lightgray}{rgb}{.9,.9,.9}
\definecolor{darkgray}{rgb}{.4,.4,.4}
\definecolor{purple}{rgb}{0.65, 0.12, 0.82}

\usepackage{float}
\usepackage{url}
\usepackage{xspace}

\usepackage{siunitx}
\usepackage[table,xcdraw]{xcolor}
\newcommand{\CFWorkers}{Cloudflare Workers\xspace}
\newcommand{\worker}{worker\xspace}

\newcommand{\mitigation}{DyPrIs\xspace}
\newcommand{\paragrabf}[1]{\noindent\textbf{#1}\ }

\usepackage{adjustbox}
\usepackage{pgfplots}
\usepackage{tcolorbox}

\newfloat{listing}{tbhp}{lst}
\floatname{listing}{Listing}

\DeclareCaptionSubType{listing}

\DeclareFloatingEnvironment[
  fileext=lob,
  listname={List of Boxes},
  name={Code Listing},
  placement=htp,
]{BOX}

\setcopyright{none}
\title{\name{}: Efficient Microarchitectural Leakage in the Cloud with Remote Timers}

\author{Martin Schwarzl}
\affiliation{%
  \institution{Cloudflare, Inc.}
  \country{USA}
}

\author{Haocheng Xiao}
\affiliation{%
  \institution{University of Edinburgh}
  \country{United Kingdom}
}

\author{Albert Pedersen}
\affiliation{%
  \institution{Cloudflare, Inc.}
  \country{USA}
}

\author{Sam Ainsworth}
\affiliation{%
  \institution{University of Edinburgh}
  \country{United Kingdom}
}

\author{Nigel Topham}
\affiliation{%
  \institution{University of Edinburgh}
  \country{United Kingdom}
}

\begin{document}

\begin{abstract}
  Edge computing solutions have become a crucial part of the industry, delivering fast, flexible and scalable applications close to the end users, with typical use cases including dynamic content creation, image resizing and chatbots.
  \CFWorkers is one such framework, which handles millions of HTTP requests per second across the world.
  To reduce start-up latency, \CFWorkers removes process-isolation boundaries between multiple tenants and leverages language-level isolation.

This architecture poses the risk of Spectre attacks. To mitigate these, \CFWorkers previously introduced several countermeasures such as restricted timer measurements, no shared memory, no multithreading and Dynamic Process Isolation (\mitigation), detecting potential attacks and process-isolating potentially malicious scripts.

In this paper, we demonstrate that the production implementation of \mitigation was insufficient.
We adopt microarchitectural amplification techniques and discover various possibilities to measure time in the production environment of \CFWorkers.
Given these techniques, we demonstrate that freezing and coarsening of timers in the security model of \CFWorkers is insufficient.
Leveraging both timing amplification and remote timers, we demonstrate a remote Spectre attack that leaks a JWT token from a co-located victim worker in the \CFWorkers production environment.
We outperform the existing attack by orders of magnitude, going from \SI{2}{\bit/\minute} to up to \SI{12}{\bit/\second} at an accuracy of \SI{99.16}{\percent}, posing an immediate risk to customer data.
Following our demonstration of the end-to-end attack, \CFWorkers mitigated it in a coordinated effort by integrating the V8 Sandbox limiting transient access to 64-bit pointers, improving the detection capabilities of \mitigation, and deploying hardware-assisted MPK-based in-process isolation to confine each tenant heap under a dedicated memory-protection key.
\end{abstract}

\maketitle




\section{Introduction}
Edge computing improves modern digital systems by moving data processing closer to users, enabling faster and smoother applications with low delays.
This enables a broad spectrum of applications ranging from dynamic content generation to real-time interactive services.
Edge-computing solutions process millions of HTTP requests per second across a globally distributed network, and the processed data contains sensitive information such as credentials, API tokens, cryptographic key material, and HTTP form data.

Edge-computing providers, such as Cloudflare Workers, use language-level isolation~\cite{CFWorkers2025Security} to reduce overheads versus process isolation, caused by long startup times and high context-switching costs~\cite{Cloudflare2025Workers,Fastly,Hyperlight}.
This results in a co-location of thousands of different tenants within the same process, meaning a single vulnerability enabling arbitrary memory reads can lead to massive data leakage~\cite{Cloudbleed,ProjectZero2017Cloudflare}.
\CFWorkers relies on the V8 runtime~\cite{Cloudflare2025Workers} to enable language-level isolation. To minimize the patch gap for memory-safety bugs, the \CFWorkers team use automated pipelines~\cite{CFWorkers2025Security} to patch V8 vulnerabilities on edge servers.


However, side-channel attacks also have to be considered in edge computing systems, where resources are shared~\cite{CFWorkers2025Security}.
In a single-process-multi-tenant systems, Spectre~\cite{Kocher2019} particularly poses a huge memory disclosure risk.
In contrast to browser vendors, who gave up on remediating in-process Spectre by introducing process isolation~\cite{Mcilroy2019Spectre,Reis2019Siteisolation}, \CFWorkers introduced several countermeasures~\cite{CFWorkers2025Security}. 
For example, the high-resolution timer available in JavaScript \texttt{performance.now} has the same coarse resolution as \texttt{Date.now}, with both returning only the time of the last I/O event, otherwise being frozen during execution.
There is also no multi-threading possible, preventing the introduction of counting threads as well as no shared memory between tenants~\cite{Schwarz2017Fantastic,CFWorkers2025Security}.

Schwarzl~\etal\cite{Schwarzl2022Robust} previously demonstrated a remote Spectre attack leaking \SI{120}{\bit/\hour},  leveraging a noisy remote timing source with a limited amount of subrequests.
The evaluated amplification technique relied on looping over the gadget with \SIx{250000} repetitions to amplify a single bit.
Moreover, to evict a single cache line, a loop over a huge array was performed, which was costly in terms of attack run-time. 
As a countermeasure, a probabilistic approach called Dynamic Process Isolation (\mitigation)~\cite{Schwarzl2022Robust} was introduced, which monitors the performance counters on a per-script base and dynamically isolates malicious-looking scripts into separate processes. 
This defense was integrated into the production system of \CFWorkers.
The main assumption of \mitigation~\cite{Schwarzl2022Robust} is that the attack is too slow to leak any sensitive data within a timeframe shorter than a measurable detection window of \SI{30}{\second}, and, therefore requires multiple executions.
Moreover, the detection assumes that all Spectre attacks form an easily distinguishable signal via performance counters compared to benign production scripts. 

Given state-of-the-art techniques to efficiently construct precise timers from microarchitectural events~\cite{leakypage,Agarwal2022Spookjs,Xiao2023Hackyracers,Kaplan2023optimization,Purnal2023Showtime} the attack has room for improvement. Still, we also identified three challenges to practicality unsolved in the previous attack~\cite{Schwarzl2022Robust}:
\begin{enumerate}
    \item C1: \textit{Limited Runtime and Co-Location of Worker Scripts.} \mitigation~\cite{Schwarzl2022Robust} assumes a co-location within a process between attacker and victim. It is unclear how to repeatedly invoke a script on the same machine and the same process.  
    \item C2: \textit{Remote Timer.} The co-located timer in \mitigation~\cite{Schwarzl2022Robust} (\cf Sec. 3.1) is assumed possible without proof, and the evaluation was performed in a local-network setup. We demonstrate how to create a stable remote timing source in the production system of \CFWorkers.
    \item C3: \textit{Robustness.} The frequent upgrades of the V8 engine in \CFWorkers require a version-agnostic proof-of-concept. Moreover, Spectre attacks are quite sensitive to systems activity i.e. noise. Thus, techniques to stabilize the attack are required with respect to the just-in time compilation, cache eviction and CPU transient-execution windows. Moreover, \mitigation is active on the production system and monitors the scripts. Thus, a successful attack requires a leak within a single or a few executions before getting detected and then process isolated.
\end{enumerate}

In this paper, we solve the above mentioned challenges and in collaboration with \CFWorkers demonstrate \textbf{a stable end-to-end remote Spectre attack}.  
For ethical reasons, the attack only aimed at our own controlled \worker scripts and no customer data has been accessed.
The attack relies on the core attack primitives such as a remote timer, stable signal amplification, a Spectre gadget and efficient memory eviction.

Leveraging WebSockets we construct a remote timer, thereby overcoming the subrequests limit for a single \worker invocation~\cite{Schwarzl2022Robust,cloudlimits}.
Our experiments show that it is possible to create a co-located remote timer to the attacker script.
We evaluate a worker-to-worker WebSocket connection leveraging existing timing source \texttt{Date.now} leading to a median resolution of \SI{1}{\milli\second}.
In addition, we analyze the resolution of a local high-resolution timer, a worker serving the local high-resolution timer via WebSocket connection, a remote timer running on an AWS instance returning high-resolution timestamps, and a timer using the Cloudflare Sandbox SDK co-located with the attacking Worker.
We demonstrate that \CFWorkers' \textit{Durable Objects} feature can be used to co-locate a high-resolution timer on the same physical machine as the attacking Worker, and that the Cloudflare Sandbox SDK enables a native \texttt{rdtsc}-based timer server on the same machine, achieving the highest accuracy of all evaluated timers.
We further demonstrate that even with a distance of over \SI{1200}{\kilo\metre}, a stable timing resolution in the hundreds of microseconds can be achieved.

We leverage a feature of \CFWorkers' \textit{Durable Objects} to circumvent the \SI{30}{\second} CPU runtime limit of a single \worker~\cite{cloudlimits}.
By using keep-alive signals, an attacker can run a single invocation of a \worker script for more than \SI{20}{\hour}. 
Moreover, this feature is helpful to achieve co-location between an attacker and victim within the same process.
The attacker compares the machine names in Cloudflare's \texttt{/cdn-cgi/trace} endpoint.
This fact undermines the design assumptions of the defenses in \mitigation~\cite{Schwarzl2021Dynamic}.
\mitigation assumes the need to average the results of multiple invocations.
Our attack leaks sensitive data already within the first invocation.
The attacker will not be process isolated during execution.
This is a fundamental limitation in the capabilities of the detection, as we show that the code paths exercised during WebSocket handling significantly increase iTLB activity, thereby suppressing the normalized ratio between branch mispredicts and iTLB hit rate, treated as a fingerprint for Spectre activity in \mitigation~\cite{Schwarzl2021Dynamic}, allowing the attack to evade detection.


To achieve robustness, we present a novel Spectre gadget that is stable across multiple V8 upgrades.
We analyze the existing techniques used to perform speculative type confusion in JavaScript.
We then enumerate typical challenges that have to be overcome such as cache eviction, memory alignment, high systems utilization and frequent V8 updates.
We leverage memory-alignment rules in V8 to enable a successful attack not reliant on cache eviction sets~\cite{Vila2019Theory,Agarwal2022Spookjs,Gruss2016Rowhammerjs}.

We evaluate our end-to-end attack in the production environment of \CFWorkers.
We demonstrate a remote Spectre attack that leaks a JWT token from a co-located victim worker.
The attack achieves a leakage rate of up to \SI{12}{\bit/\second}, outperforming previously presented attacks~\cite{Schwarzl2022Robust,Schwarzl2022Remote} by 360$\times$.
Given the demonstrated amplification techniques, the freezing and coarsening of timers in the security model of \CFWorkers is insufficient.
We conclude that the threat of in-process Spectre at \CFWorkers is higher than previously evaluated, though the \CFWorkers team found no indicators of the vulnerability being actively exploited.
Following our demonstration of the end-to-end attack, \CFWorkers mitigated it in a coordinated effort by integrating the V8 Sandbox limiting transient access to 64-bit pointers, improving the detection capabilities of \mitigation, and deploying hardware-assisted Intel-MPK-based in-process isolation to confine each tenant heap under a dedicated memory-protection key~\cite{CloudflareSandboxHardening2025}.

\textbf{Contributions.} The main contributions of this work are:
\begin{enumerate}
    \item We present techniques to co-locate two \worker scripts in production systems of \CFWorkers. 
    \item We re-evaluate remote timers in \CFWorkers, showing that high-resolution timers are in fact achievable.
    \item We identify limitations of prior Spectre gadgets under remote timers and devise methods to overcome them.
    \item We mount successful remote Spectre attacks on \CFWorkers, leaking a JWT token from a co-located victim worker at up to \SI{12}{\bit/\second}, outperforming previous attacks~\cite{Schwarzl2022Robust} by 360$\times$.
\end{enumerate}

\textbf{Outline.}
The remainder of the paper is organized as follows.
In \cref{sec:background}, we provide the necessary background. 
\Cref{sec:threat} states the threat model of our attack.
We demonstrate how we overcome the challenges when using attack primitives from Schwarzl~\etal\cite{Schwarzl2022Robust} to leak secrets from arbitrary address in~\cref{sec:attack_primitives}.
We evaluate our remote Spectre attack in \Cref{sec:evaluation}, before presenting mitigations in \Cref{sec:mitigations} and concluding in \cref{sec:conclusion}.

\section{Background and Related Work}\label{sec:background}
This section provides the necessary background for remote timing and Spectre attacks.

\subsection{Remote Timing Attacks}

To leak information about timing side channels over the internet requires \textit{remote timing}: the ability to measure execution of a process running on a remote machine. The first remote timing attacks targeted cryptographic primitives~\cite{Brumley2003Remote, Bernstein2005,Aciicmez2007d,Zhao2009cache, Jayasinghe2010remote}.
TIME~\cite{Beery2013TIME} exploits the compression ratio of encrypted data to leak HTTP cookies. 
Vanhoef~\etal\cite{VanHoef2016Heist} demonstrated that timing can be leveraged to infer the size of cross-origin resources in HTTP/2.
Van Goethem~\etal\cite{VanGoethem2020Timeless} analyzed sequential timing attacks through various data centers and presented a novel technique exploiting the response order of concurrent packets in HTTP/2 to infer the timing without measuring it. 
There were several attacks exploiting various side channels in practical settings to remotely exfiltrate data~\cite{Gruss2019page,netspectre,Kurt2020Netcat,Schwarzl2022Remote,Schwarzl2023Practical}.
Wang~\etal\cite{Wang2022Hertzbleed} leveraged DVFS to attack post-quantum cryptography. 
Kettle~\cite{Listen2024} recently demonstrated practical web timing attacks differentiating small latencies down to \SI{200}{\micro\second}.

\subsection{Spectre}
Speculative execution relies on the outcome of branch prediction to predict the targets of a branch instruction. 
If the predicted target branch turns out to be wrong, the results have to be discarded, and the correct branch is being executed.
However, this rollback does not cleanup the microarchitectural state of data that would never have been accessed under program-order correct execution, e.g. transiently accessed data still being cached, leading to Spectre attacks~\cite{Kocher2019,Canella2019A}. 
In combination with a microarchitectural side channel, the accessed data can then be recovered e.g. via cache latency.

\subsection{Spectre in JavaScript}
The typical challenge of porting Spectre to JavaScript is to craft a reliable gadget enabling transient out-of-bounds accesses.
This gadget needs to open a sufficiently large transient execution window~\cite{Roettger2021A} such that data can be both accessed and encoded into a side channel.
Given a noisy production system, measurement for a single leaked bit have to be repeatable.
Thus, another primitive is required to evict the accessed memory address at least by one cache level in order to measure whether the transient execution re-loads it or not.
The original Spectre~\cite{Kocher2019} proof-of-concept provided a JavaScript-based Spectre leaking data from an array. 
As a consequence, modern browser vendors started reducing the resolution of timers in JavaScript and suggested the use of process isolation for different origins.
Roettger and Janc~\cite{Roettger2021A} demonstrated a Spectre-PHT gadget able to address 32-bit indices relative to a \textit{TypedArray} via speculative bound checking bypass. 
At the same time, Johnson~\cite{Johnson2021Another} designed a portable Spectre exploit working on Safari and Chrome, exploiting hardware-agnostic techniques like pigeonhole eviction and memory-level parallelism to amplify a single event.
Ragab~\etal\cite{machineclear} demonstrated a transient arbitrary read primitive in Firefox by leveraging machine clears opening a transient execution window and the encoding of floating point values to construct 64-bit pointers.
Various works leverage speculative type confusion to construct a transient 64-bit read primitive ~\cite{Agarwal2022Spookjs,Schwarzl2022Robust, ILeakageKim2023}.
Wikner~\etal\cite{Wikner2022Spring} demonstrate an attack exploiting Spectre-RSB in WebAssembly to steal JWT tokens from Blazor applications in Firefox. 
Kim~\etal\cite{Kim2024Tiktag} exploited a Spectre gadget in a more recent V8 version transiently leaking ARM MTE tags. 
More recently, load value prediction and data value prediction on Apple's M2 and M3 CPUs were exploited to read sensitive data from Safari and Chrome~\cite{Kim2025Flop,Kim2025Slap}.
In all of these works, \texttt{TypedArray}s such as \texttt{Uint8Array} or \texttt{Uint32Array} are predominantly used due to their memory layout.
This layout enables both index-based memory reads and 64-bit pointer dereferences.
Both structures facilitate Spectre gadget construction in prior work and our work (\cref{sec:spectre_gadget}).

\subsection{\CFWorkers} \label{sec: background_workers}
\CFWorkers~\cite{Cloudflare2025Workers} handles millions of web requests from thousands of tenants around the world.
Its single-process design, based on V8 JavaScript isolates, allows it to run multiple thousands of tenants within one process.
This design guarantees customers' high performance demands due to less overhead when context switching between threads. 
However, this single-process design is susceptible to in-process Spectre attacks, as multiple tenants share the same virtual address space. 
A free-to-use \worker can run for up to \SI{10}{\milli\second}, and the paid version up to \SI{30}{\second}~\cite{cloudlimits}, with longer \SI{15}{\minute} run-times for Cron Trigger and Queue Consumer.
To mitigate Spectre attacks, \CFWorkers uses several countermeasures~\cite{Schwarzl2021Dynamic}.
The only available timing function \texttt{Date.now} is only updated after I/O operations are performed and machines are restarted on a daily basis~\cite{CFWorkers2025Security}. 
As a \worker is single-threaded, separate timing threads~\cite{Schwarz2017Fantastic} are not possible~\cite{CFWorkers2025Security}. 

\subsection{\mitigation}
As a countermeasure to remote Spectre attacks, Dynamic Process Isolation (\mitigation)~\cite{Schwarzl2022Robust} was introduced by Cloudflare.
The idea of \mitigation is to monitor performance counters in \worker scripts to look for signs of likely Spectre attack. 
These are normalized by the number of ITLB accesses, and based on a threshold, potentially malicious scripts are isolated into separate processes. 
Measured performance counters include branch accesses/misses, L1 accesses/misses and L3 accesses/misses~\cite{Schwarzl2022Robust}.
Schwarzl~\etal\cite{Schwarzl2022Robust} also discuss how the attack can be watered down to bypass detection by reducing L1 accesses and increasing iTLB accesses, but only at bitrates of \SI{1}{\bit/\hour}, which were thought to make the attack infeasible given a 30-second run-time limit.

\subsection{Bypassing Timing Mitigations}\label{ssec:bg:timing}

A major challenge to create a robust Spectre attack in a production system like \CFWorkers is reliable signal amplification, such that the remote timers' coarse resolution can be overcome.
Prior work has explored several approaches based on repetition, amplification gadgets, and multi-sampling.
Execution repetition~\cite{Mcilroy2019Spectre, Schwarzl2022Robust} accumulates timing differences by triggering the same leak repeatedly.
While applicable to most scenarios, the accumulation speed is slow~\cite{Mcilroy2019Spectre, Purnal2023Showtime} because most of the runtime is spent on triggering the leakage itself.
For example, Schwarzl~\etal\cite{Schwarzl2022Robust} achieved a leakage rate of only \SI{120}{\bit/\hour} with pure repetition.
Higher amplification rates are achieved by amplification gadgets.
R\"ottger and Janc~\cite{leakypage} introduced PLRU-based amplification, exploiting L1D cache replacement policies to cause many cache misses from a single state change, without repeated leakage triggers.
Hacky Racers~\cite{Xiao2023Hackyracers} generalizes this approach and proposes further amplification gadgets based on Instruction-Level Parallelism.
However, these gadgets rely on delicate microarchitectural state that is prone to invalidation by system interrupts, limiting the achievable timing difference to less than \SI{0.5}{\milli\second}.
Purnal~\etal\cite{Purnal2023Showtime} achieve timing differences up to \SI{5}{\milli\second}, and up to \SI{350}{\milli\second} with prefetch instructions.
Transient-execution-based amplification gadgets~\cite{Katzman2023gates, Kaplan2023optimization} require either native code or per-CPU fine-tuning, making them inapplicable in our setting.
In this work, we combine PLRU cache amplification~\cite{leakypage,Xiao2023Hackyracers} with nested repetition loops to overcome the interrupt-induced timing limit, achieving arbitrary amplification that is stable in production cloud environments (\cf \cref{sec:signal_amplification}).

\subsection{Cache Eviction}\label{ssec:cache-eviction}

Cache eviction is a crucial primitive to enable reliable Spectre attacks~\cite{Kocher2019,Canella2019A}.
Stable cache eviction enables repeated cache measurements on the same addresses and as well as to enhance the transient execution window due to longer delays loading data from memory~\cite{Gruss2016Rowhammerjs,Kocher2019}.
Eviction set construction relies on finding congruent memory addresses that map to the same cache set~\cite{Gruss2016Rowhammerjs}.
Vila~\cite{Vila2019Theory} developed a technique to construct precise LLC eviction sets in Chrome, which was used by Spook.js to construct a Spectre attack~\cite{Agarwal2022Spookjs}.
However, for AMD CPUs, Spook.js~\cite{Agarwal2022Spookjs} assumed a perfect LLC eviction primitive and leveraged the \texttt{clflush} instruction.
To our knowledge finding a minimal LLC eviction-set has not been achieved by any work in the literature.
Furthermore, the runtime for finding even approximately precise eviction-sets is prohibitively high for remote-timer based attacks. 
For example, Purnal~\etal\cite{Purnal2023Showtime} constructed eviction sets with a median runtime of \SI{25}{\second} in local Chrome environment with a 100$\mu\text{s}$ timer.
Schwarzl~\etal\cite{Schwarzl2022Robust} used a large array, greater than the size of the cache levels L1+L2, to evict data. 
Traversing this large array significantly slows down the attack. 
Thus, Roettger and Janc~\cite{Roettger2021A} used an eviction list to succesfully evict at least into the L2. 
This technique was further improved by Purnal~\etal\cite{Purnal2023Showtime} by using L2 congruent eviction sets.
In this work we sidestep the need for precise eviction sets by implementing the pigeonhole eviction technique, an auto-eviction technique relying on cache occupancy of L1+L2, as described by Johnson~\cite{Johnson2021Another}.




\section{Threat Model and Attack Overview}\label{sec:threat}
We assume both the attacker and the victim execute JavaScript/WebAssembly code on distinct V8 isolates in \CFWorkers, assigned to the same \worker process.
This co-location~\cite{getoffmycloud, placementvunl} assumption is reasonable due to the nature of edge computing, where the topologically nearest \worker process to the request is always picked.
Therefore, it is very likely that victim and attacker \worker are scheduled on the same machine with no additional effort\footnote{There is active data mining on the Cloudflare infrastructure on-going being able to directly access a certain machine via a Worker~\cite{cfdatamining,cfdatamininggithub}.}.
We will show that forced co-location to the same process is achievable in~\cref{sec:colocation}.
Furthermore, we assume that attacker has full control over the code running in its own isolate(s), but no native code execution. 
Thus, the attacker can only provide valid JavaScript and WebAssembly.
Moreover, there are no existing software exploits in the JavaScript Engine, nor are there any sandbox escapes. 
This threat model is identical to  Schwarzl~\etal\cite{Schwarzl2022Robust}.



\paragrabf{Attack Overview.}
\begin{figure}[t]
    \centering
    \maxsizebox{0.9\hsize}{\vsize}{
      \resizebox{\hsize}{!}{
\begin{tikzpicture}

\draw[dashed, thick, gray, rounded corners=6pt] (-1.2,-0.5) rectangle (10.2,-4.8);
\node[gray, anchor=south east, font=\normalsize] at (10.2,-4.8) {Process};

\draw[fill=red!20, rounded corners=4pt] (-0.75,-1) rectangle +(2,-3) node[pos=.5] {\parbox{1.75cm}{\centering Attacker}};

\draw[fill=orange!20, rounded corners=4pt] (7,-1) rectangle +(2,-3) node[pos=.5] {\parbox{1.75cm}{\centering Worker Secret}};

\draw[fill=green!20, rounded corners=4pt] (4,4) rectangle +(4,-1.5) node [pos=.5] {\parbox{3.5cm}{\centering \large Remote Timer}};

\draw[->,>=stealth,thick] (-0.5,-1) to node[midway,sloped,above] {\Large \textcircled{\raisebox{-1.1pt}{1}} Start measurement}(4,2.45);

\draw[->,>=stealth,thick] (1.25,-2) to node[midway,above] {\Large \textcircled{\raisebox{-1.1pt}{2}} Transiently access \textbf{bit}}(7,-2);

\node at (8,-4.3) {\Large \textcircled{\raisebox{-1.1pt}{3}} Amplify signal};

\draw[->,>=stealth,thick] (6,2.5) to node[midway,sloped,above] {\Large \textcircled{\raisebox{-1.1pt}{4}} Measure RTT}(1.25,-1);

\end{tikzpicture}
}
    }
    \caption{High-level overview of a remote Spectre attack on \CFWorkers.}
    \label{fig:attack_overview}
\end{figure}
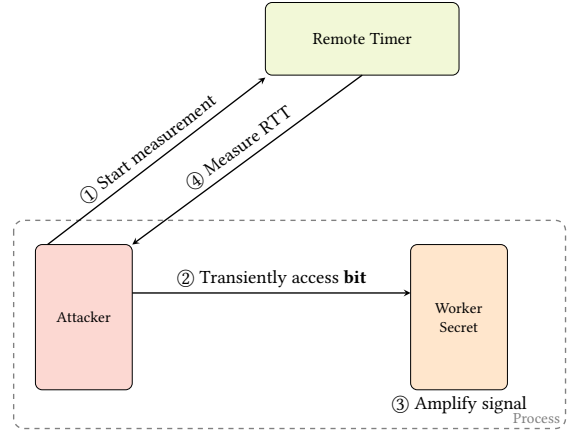
Both the attacker and victim script are co-located within the same process and have a closely co-located remote timer.
The presented attack targets the production instances of \CFWorkers, registered via Cloudflare's dashboard.
Therefore, all active mitigations and detections are in place.
The attack's success rate may be influenced by other users systems activity such as cache activity and network contention from other tenants and processes.

\cref{fig:attack_overview} illustrates an overview of a remote side channel attack on \CFWorkers consisting of four steps, which is executed repeatedly on each bit of data.
\textcircled{\raisebox{-1.1pt}{1}} The attacker initiates a connection to a remote timer and performs a request to record a timestamp.
\textcircled{\raisebox{-1.1pt}{2}} The attacker leverages a JavaScript Spectre gadget to transiently access arbitrary data. Multiple gadgets may be utilized through the attack process to finally leak the targeted secret data.
\textcircled{\raisebox{-1.1pt}{3}} The attack uses amplification techniques to make the bit signal resilient to network and system noise.
\textcircled{\raisebox{-1.1pt}{4}} A second request is sent to the remote timing server to generate another timestamp.
Based on the delta of the two timestamps, the attacker infers the bit value.

\section{Attack Primitives}\label{sec:attack_primitives}
Here we describe our main attack primitives, which solve the above stated challenges of process co-location in~\cref{sec:colocation}, remote timers in~\cref{sec:remote-timer}, and robust timing-signal amplification in~\cref{sec:signal_amplification}, before describing a selection of practical facilitating Spectre setup and execution gadgets in \cref{sec:spectre_gadget} to pull off a full end-to-end attack.

\subsection{Bypassing Time Limits and Achieving Co-Location}\label{sec:colocation}

Co-location of attacker and victim on the same physical host has been studied in IaaS cloud environments~\cite{getoffmycloud,placementvunl}, but the serverless model used in \CFWorkers\ introduces different constraints: tenants share a single process rather than a VM, and execution is bounded by per-invocation CPU time limits.
Prior work~\cite{Schwarzl2022Robust} assumed that an attacker and victim worker can be co-located within the same process, and that the attacker can sustain execution beyond the per-invocation time limit, but left both claims unproven in production.
We demonstrate both for the first time in the production system of \CFWorkers.

\paragrabf{Durable Objects.}
Durable Objects~\cite{Cloudflare2025DO} are special Workers that run in only one place at any given time.
Their main purpose is creating long-living and stateful applications, e.g. a chat.
This allows them to be used as a synchronization primitive and as message broker between multiple Workers.
They are designed to stay alive indefinitely with regular invocations.
Standard Workers are limited to 30 seconds of CPU time per invocation, which limits the amount of data that can be leaked per invocation, as attack setup must be repeated each time a Worker initiates.
While these steps can be repeated they add a constant overhead to the attack and reduce the leakage rate.

\paragrabf{Keeping a Durable Object alive.}
Like a standard \worker script, Durable Objects are also limited to \SI{30}{\s} of CPU time. However, for a Durable Object this limit is reset to \SI{30}{\s} each time an HTTP request is received. 
We can leverage this fact to significantly extend the CPU time using WebSocket messages, which are internally treated like HTTP requests.
Therefore, to keep a Durable Object alive, one simply has to initiate a WebSocket connection to the Durable Object and send a message at least every 30 seconds.
Since Durable Objects are still single-threaded, the Worker must regularly wait for I/O for any WebSocket messages to be received. 
If the thread is continuously blocked for more than 30 seconds, the Worker does not get a chance to process the incoming WebSocket message. 
We craft an experiment that constantly sends a keep-alive signal to the Durable Object script.
We observe  that the isolate can be kept alive at least for \SI{5}{\hour} up to more than \SI{20}{\hour}. 
This clearly bypasses the 30-second limit.

\paragrabf{Identifying a physical machine.}
Every domain being added to Cloudflare receives a custom tracing endpoint \texttt{/cdn-cgi/trace}.
This endpoint contains various interesting data such as TLS information, the HTTP version, data centre name, a timestamp value and the \textbf{fl} value naming a data centre e.g. \textbf{fl=114f227}, whereas 114 is the id of the data centre and 227 the id of the machine. 
This endpoint has been leveraged for data-mining to learn about the infrastructure of Cloudflare~\cite{cfdatamining}.

\paragrabf{Hitting the same process.}
Any memory address whose value remains stable over time can serve as a fingerprint to identify a running process or isolate. 
Using a Spectre gadget that can read an arbitrary 64-bit location, an attacker can probe such addresses and compare the observed values to determine whether the same isolate or process is still running. 
Examples of useful targets include process-wide structures or globals like the isolate root and fixed-address mappings such as the vDSO. 
An attacker may first leak a candidate fixed address, then repeatedly probe that address to verify the target process. 
Once the attacker has identified the target process, Durable Objects can be used to force an isolate onto that process for further exploitation.

\begin{tcolorbox}[boxsep=1pt,left=2pt,right=2pt,top=1pt,bottom=1pt]
  \textbf{C1: Limited Runtime and Co-location}
  We use Durable Objects to keep a script running in the same isolate for hours. The \texttt{/cdn-cgi/trace} endpoint identifies the target machine. We then spawn Durable Objects until two isolates share one process. 
\end{tcolorbox}

\subsection{Remote Timer on Cloudflare Workers}\label{sec:remote-timer}
\CFWorkers freezes the time during raw CPU execution and only updates the timers based on IO events, e.g. performing a requests with fetch API or an update of Workers KV.
Prior work~\cite{Schwarzl2022Robust} used subrequests as a timing proxy, treating each outbound fetch as a coarse clock tick.
This approach is constrained by the \CFWorkers subrequest limit~\cite{cloudlimits} and only demonstrated in a local-network setup, leaving its viability in the production environment unproven.
We instead leverage the WebSocket API, which is not subject to the subrequest limit, to construct a high-frequency remote timer that operates entirely within the production environment.
We enumerate and compare the realizations of the method below.

\paragrabf{WebSocket Remote Timer.}
\CFWorkers offers the WebSocket API to communicate with remote WebSocket servers.
This interface allows it to open a WebSocket connection.
Note that the communication overhead introduced by the WebSocket connection does not count towards the actual CPU time of a Worker instance.
Instead, it is part of the Wall time.
We verified this in the Cloudflare dashboard, as it reflects statistics about the CPU and Wall time.
In contrast to the amount of subrequests, \CFWorkers does not restrict the amount of messages being transmitted in WebSocket connections. 
For many subsequent WebSocket calls the CPU time barely increased. 
Thus, for the paid version with an offered runtime of \SI{30}{\second} per Worker invocation, the Wall time can be several minutes of execution time.

Our WebSocket communication uses one call to \textbf{mark} a timestamp. 
To receive the delta, another command is used to compute the difference between two marked timestamps. \Cref{lst:timerJS} illustrates the usage of the remote timer and how to trace an event.
This approach can also be optimized by collecting the timestamps for a certain event on the server first and then gathering the list of results directly from the server.

\begin{listing}[t]
    \begin{lstlisting}[caption={Interface (client) used to communicate with a remote WebSocket server.},label={lst:timerJS},language=JavaScript,style=customjs]
const timer = new RemoteTimer("rt.com");
timer.mark("mark S1");
event();
timer.mark("mark E1");
const delta = await timer.delta("delta S1 E1");
    \end{lstlisting}
    \vspace{-0.6cm}
\end{listing}

\begin{listing}[t]
    \begin{lstlisting}[caption={Remote Websocket timer as \worker script.},label={lst:timerServer},language=JavaScript,style=customjs]
async fetch() {
  server.addEventListener('message', (event: MessageEvent) => {
    const now = Date.now();
    const message = JSON.parse(event.data);
    switch(message.type) {
      case 'mark':
        marks.set(message.name, now); break
      case 'delta':
        server.send(marks.get(message.marks[0])-
        marks.get(message.marks[1]))
      }
    })
}
    \end{lstlisting}
    \vspace{-0.6cm}
\end{listing}

\paragrabf{Co-Located Worker as Timer}
As \texttt{Date.now} is updated on I/O operations, we analyze and exploit the available API calls that can trigger timer updates.
The returned resolution of \texttt{Date.now} is still in the millisecond range\footnote{https://developers.cloudflare.com/workers/runtime-apis/performance/}. 
We leverage \texttt{scheduler.wait}, to trigger a short-lived I/O operation. \texttt{Date.now} then updates its value even though the I/O operation is short.
Using the interface described in \Cref{lst:timerJS}, we spawn another Worker script that acts as a WebSocket server, which calls \texttt{Date.now} for each mark command and provides the delta between two timestamps (\textbf{WS+Date.now}).
\Cref{lst:timerServer} illustrates a worker script acting as remote WebSocket timer.
\paragrabf{External WebSocket Server as Timer.}
To reduce latency, \CFWorkers routes requests to the topologically closest data center.
We host a dedicated WebSocket server on an AWS EC2 instance in eu-north-1 (Stockholm), approximately \SI{1200}{\kilo\meter} from the Workers data center (\textbf{WS+AWS}).
This server uses the \texttt{rdtsc} instruction for high-resolution timestamps and is built with the uWebSocket C++ library\footnote{\url{https://github.com/uNetworking/uWebSockets/}} with \texttt{TCP\_NODELAY} and \texttt{TCP\_QUICKACK} enabled.
We deliberately chose this distance to demonstrate that a practical timer remains achievable even over a wide-area connection.

\paragrabf{Co-Located Durable Object as Timer.}
As discussed in~\cref{sec:colocation}, Durable Objects co-locate two isolates in the same process.
The specific process does not matter.
The attacker invokes the victim's script through an HTTP request.
This request places the victim in the attacker's process.
At the time of writing, Cloudflare runs a single \texttt{workerd} process per machine.
Therefore, the attacker only needs to co-locate two isolates on one machine.
The attacker reads the current machine identifier from the \texttt{/cdn-cgi/trace} endpoint.
The attacker spawns new Durable Objects, where each one starts a new isolate.
The attacker repeats this until two isolates share the same process.
Following the birthday paradox~\cite{cormen2022introduction}, this search completes in about one second.
The co-located Durable Object then serves \texttt{Date.now} timestamps over a WebSocket (\textbf{WS+DO}).

\paragrabf{Cloudflare Sandbox as Timer.}
The Cloudflare Sandbox SDK~\cite{CFSandbox} runs an arbitrary containerized process directly on Cloudflare infrastructure.
Cloudflare offers it on paid Workers plans.
Each Sandbox requires a Durable Object that hosts the container on the same machine.
Co-locating the Sandbox therefore reduces to the Durable Object co-location of~\cref{sec:colocation}.
The container runs native code and serves \texttt{rdtsc} timestamps over a WebSocket (\textbf{WS+Sbx}).
This timer removes the off-platform network hop.

\paragrabf{Timer Resolution.}
To measure the effective resolution of our remote timers, we conduct the following experiment.
We use an internal binding to generate controlled timing delays.
Only the \CFWorkers team can assign this binding.
It provides a high-resolution timer.
Our experiment compares five remote timers, each connected through a WebSocket.
We measure their resolution for delays from \SI{10}{\micro\second} to \SI{2}{\milli\second}.
The \textbf{WS+Date.now} timer returns \texttt{Date.now}.
The \textbf{WS+PerfNow} timer returns the internal high-resolution timer value.
Attackers cannot assign this binding, so we include \textbf{WS+PerfNow} only as the reference upper bound for a co-located remote timer.
The \textbf{WS+AWS} timer runs on an AWS instance approximately \SI{1200}{\kilo\meter} from the Worker.
The \textbf{WS+DO} timer runs inside a co-located Durable Object.
It uses \texttt{Date.now}, as a Durable Object exposes no high-resolution clock.
The \textbf{WS+Sbx} timer runs inside the co-located Sandbox introduced above.
Using the internal high-resolution timer, we separately measure the ground truth of each delay.

Within the same Worker invocation, we perform \SIx{250} measurements per remote timer.
We repeat the entire experiment \SIx{20} times.
In total, we collect \SIx{5000} samples per remote timer.
We compute the median and the standard deviation over all samples per delay.
We observe variances on the order of several hundred microseconds across the timers.
\Cref{tab:medianvals} summarizes the median values for all timer types.
The \textbf{WS+Date.now} timer reaches an effective resolution of roughly \SI{1}{\milli\second}.
The co-located \textbf{WS+DO} timer reaches the same \SI{1}{\milli\second} resolution.
A Durable Object exposes only \texttt{Date.now}, so co-location alone does not improve the resolution.
The \textbf{WS+PerfNow}, \textbf{WS+AWS}, and \textbf{WS+Sbx} timers reach sub-millisecond median resolution.
\textbf{WS+PerfNow} is not attacker-deployable, and \textbf{WS+AWS} pays a wide-area network hop.
Only \textbf{WS+Sbx} combines sub-millisecond resolution, co-location, and attacker deployability.
This motivates the native \texttt{rdtsc} timer inside the co-located Sandbox instead of a Durable Object.
The median resolution alone does not determine whether a timer is usable in practice.
All remote timers are subject to system workload.
Their standard deviations range from several hundred microseconds up to a few milliseconds and depend on the current system utilization (\cref{fig:timeseries_remote_timer}).
The \textbf{WS+Sbx} timer reaches the lowest median error.
No remote timer therefore resolves an individual sub-millisecond delay reliably.
We address this variance through signal amplification (\cref{sec:signal_amplification}).
We amplify the difference between a cache hit and a cache miss up to the millisecond scale, where it exceeds the timer standard deviation.
At this scale, all five remote timers separate cache hits from cache misses reliably.
We further evaluate the effectiveness of our remote timers in~\Cref{sec:evaluation}, where we observe that for a sufficiently large signal, \SIx{5} to \SIx{40} samples are sufficient to differentiate between a leaked '0' and '1' bit.

\begin{table}[h!]
  \centering
  \footnotesize
  \setlength{\tabcolsep}{4pt}
  \begin{tabular}{l|rrrrrrr}
  \toprule
  \textbf{Timer} & \multicolumn{7}{c}{\textbf{Target delay [ms]}} \\
  & 0.01 & 0.05 & 0.10 & 0.25 & 0.5 & 1.0 & 2.0 \\
  \midrule
  WS+Date.now & 0.000 & 0.000 & 0.000 & 0.000 & 1.000 & 1.000 & 2.000 \\
  WS+PerfNow  & 0.071 & 0.112 & 0.162 & 0.313 & 0.586 & 1.096 & 2.101 \\
  WS+AWS      & 0.032 & 0.074 & 0.132 & 0.281 & 0.568 & 1.091 & 2.122 \\
  WS+DO       & 0.000 & 0.000 & 0.000 & 0.000 & 1.000 & 1.000 & 2.000 \\
  WS+Sbx      & 0.031 & 0.057 & 0.116 & 0.292 & 0.573 & 1.100 & 2.125 \\
  Local       & 0.010 & 0.050 & 0.100 & 0.250 & 0.500 & 1.000 & 2.001 \\
  \midrule
  \multicolumn{8}{l}{\textit{Standard deviation $\sigma$ [ms]}} \\[2pt]
  WS+Date.now & 0.320 & 0.332 & 0.449 & 0.490 & 0.503 & 0.334 & 0.743 \\
  WS+PerfNow  & 0.079 & 0.079 & 0.064 & 0.702 & 0.241 & 0.262 & 0.518 \\
  WS+AWS      & 0.137 & 0.145 & 0.168 & 0.174 & 0.317 & 0.446 & 0.675 \\
  WS+DO       & 0.000 & 0.196 & 0.300 & 0.455 & 0.499 & 0.481 & 0.472 \\
  WS+Sbx      & 0.163 & 0.109 & 0.133 & 0.187 & 0.231 & 0.259 & 0.319 \\
  Local       & 0.038 & 0.023 & 0.014 & 0.013 & 0.018 & 0.007 & 0.010 \\
  \bottomrule
  \end{tabular}
  \caption{Median timing and std-dev for the evaluated timers (in milliseconds), measured against controlled target delays from \SI{10}{\micro\second} to \SI{2}{\milli\second}. Each cell aggregates \SIx{5000} samples.}
  \label{tab:medianvals}
\end{table}

\begin{figure}[t]
    \centering
    \begin{subfigure}{\hsize}%
      \begin{tikzpicture}
    \begin{axis}[
        style={font=\footnotesize},
        xlabel={Network Timer [ms]},
        ylabel={Density},
        width=\hsize,
        height=3.5cm,
        ymin=0,
        axis x line=bottom,
        axis y line=left,
        axis background/.style={fill=none},
        ]
        \addplot[red, solid, thick] table [x index=0, y index=1, col sep=comma] {data/hist_hits.csv};
        \addplot[blue, dashed, thick] table [x index=0, y index=1, col sep=comma] {data/hist_misses.csv};
        \draw[red, solid, line width=1.5pt] ({axis cs:1.44,0}) -- ({axis cs:1.44,100});
        \draw[blue, dashed, line width=1.5pt] ({axis cs:1.03,0}) -- ({axis cs:1.03,100});
    \end{axis}
\end{tikzpicture}%
    \end{subfigure}
    \begin{subfigure}{\hsize}%
      \begin{tikzpicture}
    \begin{axis}[
        style={font=\footnotesize},
        xlabel={Ground Truth [ms]},
        ylabel={Density},
        width=\hsize,
        height=3.5cm,
        ymin=0,
        axis x line=bottom,
        axis y line=left,
        axis background/.style={fill=none},
        ]
        \addplot[red, solid, thick] table [x index=0, y index=1, col sep=comma] {data/hist_gt_hits.csv};
        \addplot[blue, dashed, thick] table [x index=0, y index=1, col sep=comma] {data/hist_gt_misses.csv};
        \draw[red, solid, line width=1.5pt] ({axis cs:1.44,0}) -- ({axis cs:1.44,100});
        \draw[blue, dashed, line width=1.5pt] ({axis cs:1.03,0}) -- ({axis cs:1.03,100});
    \end{axis}
\end{tikzpicture}%
    \end{subfigure}
    \vspace{-0.35cm}
    \caption{Kernel-density estimate of \SIx{50} amplified cache-hit (solid) and cache-miss (dashed) timing measurements. Top: network timer showing overlap due to jitter. Bottom: ground truth with
     cleaner separation. Vertical lines mark respective medians.}
    \label{fig:kde_remote_timer_gt}
\end{figure}
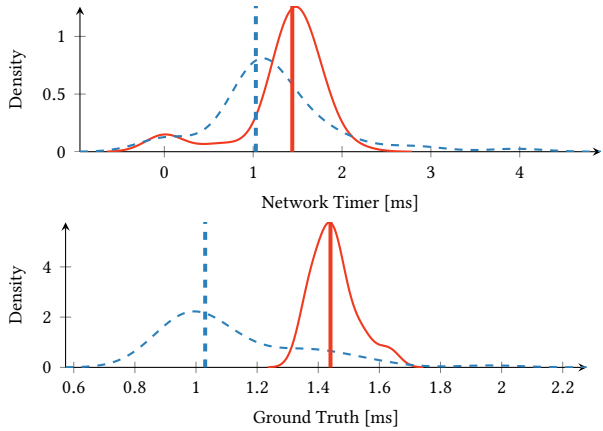

\paragrabf{Resolution of amplified delta.}
To further measure the stability of the timers, we conduct an experiment that creates a \SI{400}{\micro\second} delay between an amplified cache hit and miss using the PLRU gadget from \texttt{leaky.page}~\cite{leakypage}.
We plot the timing distributions as kernel-density estimates.
As shown in~\cref{fig:kde_remote_timer_gt}, the two distributions are visually distinguishable even from \SIx{50} randomly drawn measurements, confirming that the timing signal is strong enough to classify cache hits and misses by eye.
These results are consistent with remote timing results in similar settings~\cite{Schwarzl2022Remote,Schwarzl2023Practical,Listen2024}.

\begin{tcolorbox}[boxsep=1pt,left=2pt,right=2pt,top=1pt,bottom=1pt]
  \textbf{C2: Remote Timer.}
  Despite \CFWorkers freezing timers during CPU execution, our co-located Sandbox timer (\textbf{WS+Sbx}) achieves sub-millisecond median resolution (\cref{tab:medianvals}).
  Individual measurements exhibit standard deviations of \SI{0.5}{\milli\second} to a few milliseconds under production noise.
  Amplifying the signal to the millisecond scale (\cref{sec:signal_amplification}) makes all five timers sufficient to distinguish cache hits from misses.
\end{tcolorbox}

\subsection{Signal Amplification}\label{sec:signal_amplification}

A challenge for remote Spectre attacks is bridging the gap between the nanosecond-scale timing differences produced by cache side channels and the millisecond-scale resolution of remote timers.
Prior work~\cite{Schwarzl2022Robust} relied on pure repetition, executing the Spectre gadget hundreds of thousands of times to accumulate a measurable signal.
This achieved leakage of only \SI{120}{\bit/\hour}, as the majority of run-time was consumed by repeated gadget setup rather than actual signal accumulation.

\begin{figure}[t]
  \centering
  \begin{subfigure}{0.25\hsize}%
  \begin{tikzpicture}[scale=0.65, every node/.style={scale=0.65}]
\useasboundingbox (-1.5,-9) rectangle (4,1); 
    \tikzstyle{hit}=[fill=green!50, draw=black, rectangle, minimum width=2cm, minimum height=0.4cm]
    \tikzstyle{miss}=[fill=red!50, draw=black, rectangle, minimum width=2cm, minimum height=0.9cm]  
    \tikzstyle{interrupt}=[fill=blue!50, draw=black, rectangle, minimum width=2cm, minimum height=0.55cm]
    \tikzstyle{setup}=[fill=gray!30, draw=black, rectangle, minimum width=2cm, minimum height=0.55cm]
    
    \node at (-2, 1) {Leak Bit 0};
    \node at (2, 1) {Leak Bit 1};
    
    \draw[->] (-3.5, 1) -- (-3.5, -8.7) node[below] {Time};

    \node[setup] (b1) at (-2, 0) {Setup};
    \node[hit] (b2) at (-2, -0.65) {L1 Hit};
    \node[hit] (b3) at (-2, -1.25) {L1 Hit};
    \node[hit] (b4) at (-2, -1.85) {L1 Hit};
    \node[hit] (b5) at (-2, -2.45) {L1 Hit};
    \node at (-1.8, -3.4) {\huge$\vdots$};  
    
    \node[interrupt] (b6) at (-2, -4.4) {Interrupt};
    
    \node[hit] (b7) at (-2, -5.2) {L1 Hit};
    \node[hit] (b8) at (-2, -5.8) {L1 Hit};
    \node[hit] (b9) at (-2, -6.4) {L1 Hit};

    \node at (-2, -7) {\huge$\vdots$};  

    \node[setup] (a1) at (2, 0) {Setup};
    \node[hit] (a2) at (2, -0.65) {L1 Hit};
    \node[miss] (a3) at (2, -1.5) {L1 Miss};  
    \node[hit] (a4) at (2, -2.35) {L1 Hit};
    \node[miss] (a5) at (2, -3.2) {L1 Miss};  
    \node at (2.2, -4.4) {\huge$\vdots$};  
    
    \node[interrupt] (a6) at (2, -5.4) {Interrupt};
    
    \node[hit] (a7) at (2, -6.2) {L1 Hit};
    \node[hit] (a8) at (2, -6.8) {L1 Hit};
    \node[hit] (a9) at (2, -7.4) {L1 Hit};

    \node at (2, -8) {\huge$\vdots$};  

    \draw[->, dashed, red, thick] (2, -8.5) to[out=270, in=270] (3.3, -8.5) 
        to[out=90, in=-90] (3.3, 0.3) to[out=90, in=90] (2, .3);

    \draw[->] (b1.south) -- (b2.north);
    \draw[->] (b2.south) -- (b3.north);
    \draw[->] (b3.south) -- (b4.north);
    \draw[->] (b4.south) -- (b5.north);
    \draw[->] (b5.south) -- (b6.north);
    \draw[->] (b6.south) -- (b7.north);
    \draw[->] (b7.south) -- (b8.north);
    \draw[->] (b8.south) -- (b9.north);

    \draw[->] (a1.south) -- (a2.north);
    \draw[->] (a2.south) -- (a3.north);
    \draw[->] (a3.south) -- (a4.north);
    \draw[->] (a4.south) -- (a5.north);
    \draw[->] (a5.south) -- (a6.north);
    \draw[->] (a6.south) -- (a7.north);
    \draw[->] (a7.south) -- (a8.north);
    \draw[->] (a8.south) -- (a9.north);
\end{tikzpicture}%
  \end{subfigure}
  \caption{Timing-difference accumulation for leaking bit 0 vs.\ bit 1. The setup stage prepares the PLRU cache state and encodes the transiently accessed secret. Before an interrupt, PLRU accumulates timing difference through cache hit/miss asymmetry. If an interrupt destroys cache state, the PLRU gadget stops accumulating. With nested repetition (red dashed line), each iteration re-establishes the state to continue accumulating.}
  \label{fig:timeline}
\end{figure}
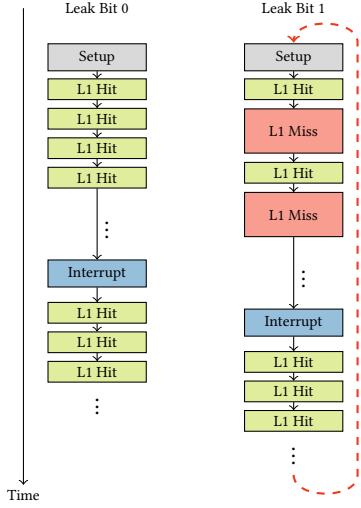

PLRU-based amplification~\cite{leakypage} improves upon this by exploiting the widely deployed pseudo-last-recently-used L1D cache replacement policy to generate many cache misses from a single leaked bit, achieving higher amplification rates without repeated leakage triggers.
However, PLRU amplification alone is limited to approximately \SI{400}{\micro\second} of timing difference before system interrupts and context switches destroy the cache state.
A \SI{400}{\micro\second} difference is sufficient under low timer noise.
Under production noise, however, the remote-timer standard deviation reaches up to a few milliseconds (\cref{tab:medianvals}) and can exceed this difference.
This motivates a further amplification step, \cf \cref{fig:timeline}.

More powerful amplification approaches from the literature are not directly applicable in our setting.
Showtime~\cite{Purnal2023Showtime} and Hacky Racers~\cite{Xiao2023Hackyracers} rely on performance counter access or fine-grained ILP contention control that is unavailable inside the \CFWorkers JavaScript sandbox.
Gates of Time~\cite{Katzman2023gates} requires a co-resident attacker process capable of manipulating contention across hardware threads, which the serverless model does not permit.
We therefore choose and extend PLRU-based amplification as it relies solely on the L1D cache replacement policy, which is uniformly implemented on all AMD EPYC CPUs deployed by \CFWorkers, and requires no privileged access or cross-process coordination.

\paragrabf{Nested-Loop Amplification.}
We overcome the \SI{400}{\micro\second} interrupt-induced ceiling of standalone PLRU amplification by combining it with an outer repetition loop, creating a nested-loop structure that achieves \emph{arbitrary} amplification independent of interrupt frequency.
\Cref{fig:timeline} illustrates the approach: rather than attempting to amplify a single transient execution indefinitely, we repeat the full sequence of setup, transient leak, and PLRU amplification across many iterations.
Each iteration independently encodes the leaked bit into cache state and amplifies it.
If an interrupt destroys the L1 state during one iteration, the subsequent iteration re-establishes the state and continues accumulating timing difference.
\Cref{lst:fineGrainTimer} shows the nested-loop structure.
The inner loop (\texttt{INNER\_REP}) controls the PLRU amplification within a single iteration, while the outer loop (\texttt{OUTER\_REP\_NUM}) repeats the full gadget execution.
A remote timer measurement brackets each outer-loop block, and we take the median over multiple samples (\texttt{SAMPLE\_NUM}) for robustness.
The total accumulated timing scales linearly with the number of inner-loop repetitions.
The amplification factor is therefore a freely tunable parameter, \cf \cref{fig:amplification_scaling} in \Cref{app:amplification}.
This approach has two key properties.
First, the timing difference is \emph{unbounded}. By increasing the number of outer-loop iterations, we can produce arbitrarily large timing differences that exceed any practical remote timer resolution.
Second, the approach is \emph{resilient to noise}. If an interrupt destroys L1-cache state mid-iteration, the PLRU oscillation loop produces only hits for both 0 and 1 cases, so the disrupted iteration adds equal time to both paths so does not corrupt accumulated signal.
\paragrabf{PLRU Oscillation Gadget.}
The PLRU amplification is implemented in WebAssembly to avoid JIT-induced nondeterminism in the oscillation loop.
Each invocation proceeds in three phases.
First, a \textit{clearSet phase} loads eight cache-line-aligned addresses congruent to the target cache set, evicting all prior occupants.
Second, a \textit{prime phase} loads seven eviction entries plus a \textit{keepAlive} entry, fully occupying the 8-way set in a defined PLRU tree state.
Third, the \textit{oscillation loop} accesses all seven eviction entries interleaved with repeated accesses to \textit{keepAlive}.
Because \textit{keepAlive} is touched after every entry, it consistently holds the most-recently-used position and is never evicted.
Whether the secret-dependent cache line displaced one of the seven entries during the transient execution determines whether those entries hit or miss during oscillation, producing the timing asymmetry that encodes the leaked bit.
Our approach uses the \texttt{oscillateTreePLRU2}~\cite{leakypage} that accesses the \textit{keepAlive} after every entry. 
The increase in \textit{keepAlive} frequency makes the PLRU tree state resilient to incidental cache accesses from JIT-compiled code and OS interrupts, which are more frequent in the \CFWorkers execution environment than in a local browser.
The resulting reduction in absolute timing delta per iteration is compensated by the outer repetition loop in our nested-loop amplification (\cref{sec:signal_amplification}).
\paragrabf{Timer Calibration.}
Before leaking secrets, we calibrate the timing threshold dynamically for each worker invocation.
We leak multiple known 0 and 1 bits and verify that the resulting timing distributions are well separated.
The classification threshold is set as the midpoint between the median timing values for 0 and 1 bits.
This per-invocation calibration accounts for varying runtime conditions across different requests.

\begin{tcolorbox}[boxsep=1pt,left=2pt,right=2pt,top=1pt,bottom=1pt]
  \textbf{C3.1: Robustness (Signal Amplification).}
  Existing amplification techniques do not transfer to our setting. PLRU amplification alone is limited by interrupt frequency and pure repetition is infeasible for practical attacks. Our nested-loop approach combines both to achieve arbitrary, interrupt-resilient amplification compatible with remote timers.
\end{tcolorbox}

\begin{listing}[t]
    \begin{lstlisting}[caption={Nested-loop PLRU amplification with remote timer.}, label={lst:fineGrainTimer}, language=JavaScript, style=customjs]
for (let s = 0; s < SAMPLE_NUM; s++) {
    timer.mark("mark S"+s);
    for (let r = 0; r < OUTER_REP_NUM; r++) {
        setup(); // branch mistrain, cache control
        leak(secret_bit); // transient read
        PLRU(cacheset, INNER_REP); // amplify
    }
    timer.mark("mark E"+s);
}
delta_t = fetchFromServer(SAMPLE_NUM);
return median(delta_t);\end{lstlisting}
    \vspace{-0.6cm}
\end{listing}


\subsection{Spectre Gadgets}\label{sec:spectre_gadget}

The nested-loop amplification of \cref{sec:signal_amplification} wraps the \texttt{setup} and \texttt{leak} calls of a Spectre gadget.
We now describe the gadgets that realize this transient read.
Following prior work~\cite{Agarwal2022Spookjs, Schwarzl2022Robust}, our attack chain requires two types of Spectre gadgets:
a \textit{heap-address gadget} that leaks address information from objects allocated in proximity to attacker-controlled objects, and
a \textit{secret-leakage gadget} that reads from arbitrary 64-bit addresses.
The heap-address gadget provides the address anchor required to bootstrap the secret-leakage gadget, which then extracts data from victim workers' address space.
We enhance prior heap-address gadgets~\cite{Schwarzl2022Robust, Agarwal2022Spookjs} to eliminate the fragile cacheline-alignment requirement, and integrate pigeonhole-style eviction~\cite{Johnson2021Another} to remove explicit eviction-set construction entirely.
We also simplify prior secret-leakage gadgets~\cite{Schwarzl2022Robust, Agarwal2022Spookjs} by eliminating the fake-pointer setup stage, by utilizing allocator object co-location and details of V8 \texttt{Uint8Array} backing buffers.

\paragrabf{Heap-Address Gadget.} The heap-address gadget leaks address information from V8 heap objects allocated near attacker-controlled objects.
This information serves as the address anchor for the subsequent secret-leakage gadget.

\begin{figure}[t]
  \centering
  \includegraphics[trim=100 150 200 0, clip, width=\hsize]{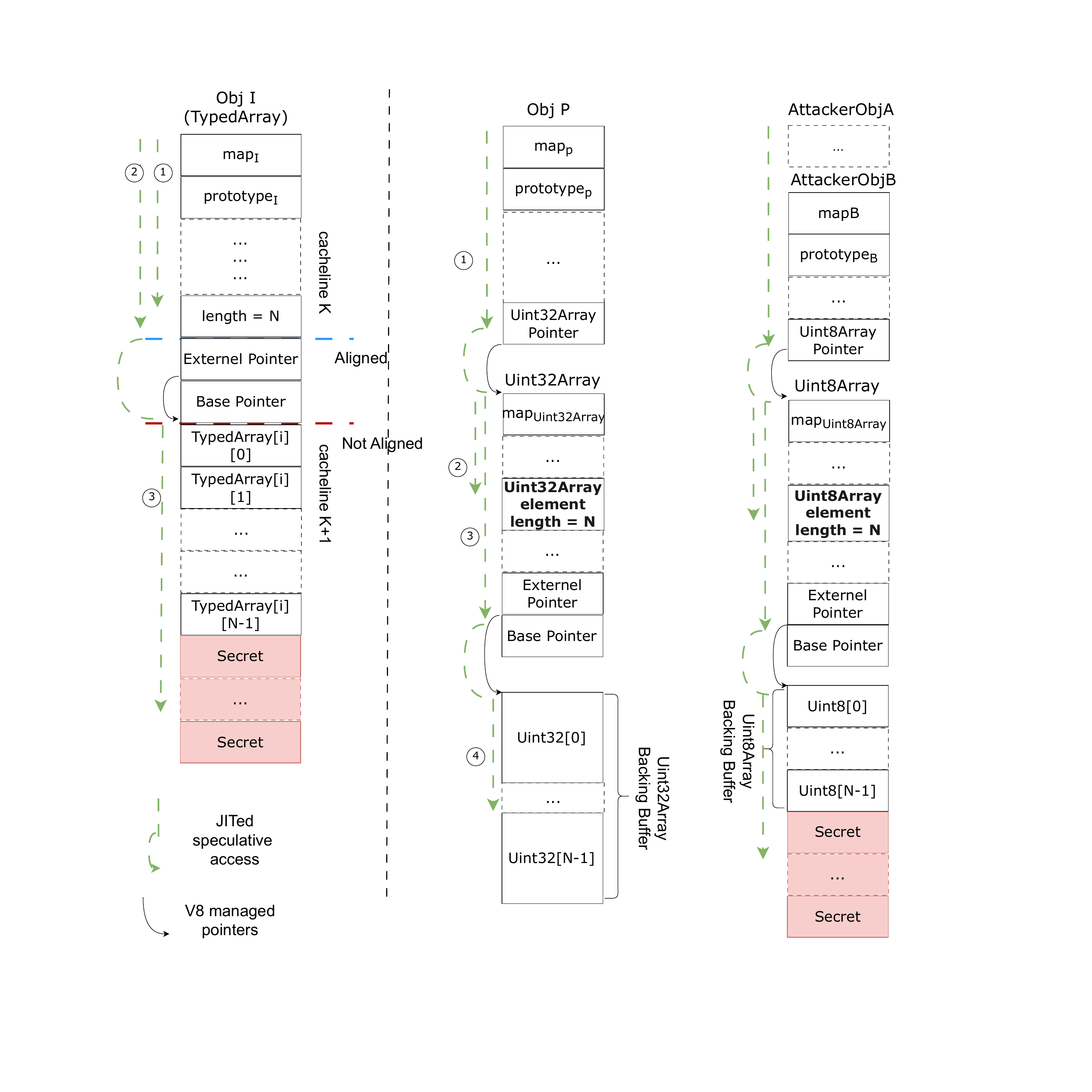}
  \vspace{0cm}
  \caption{Memory layout for our heap-address gadget. The speculative type confusion between \texttt{Uint8Array} and \texttt{Uint32Array} allows reading past the intended boundary to leak adjacent address information.}
  \label{fig:leakheap}
\end{figure}

Prior heap-address gadgets~\cite{Schwarzl2022Robust, Agarwal2022Spookjs} rely on a speculative out-of-bounds access that requires specific cacheline-level alignment: the length field and the buffer pointer of a \texttt{TypedArray} must be split across two cache lines so that one can be evicted without the other.
Finding this alignment requires allocating objects across approximately \SI{2}{\mebi\byte} of heap space and is brittle across V8 versions and allocator changes.
Additionally, these gadgets require explicit eviction-set construction before each invocation: in our measurements, finding a working eviction set requires between 1 and 50 attempts and takes approximately \SI{15}{\second}, with a resulting \SI{50}{\percent} per-gadget failure rate that significantly reduces end-to-end leakage throughput.
We replace the alignment-sensitive out-of-bounds approach with a speculative type-confusion~\cite{kirzner2021analysis} gadget.
Under speculative execution, the gadget accesses a \texttt{Uint8Array} object as if it were a \texttt{Uint32Array}, reading 4~bytes per access instead of 1.
This mismatch allows the transient path to read past the intended boundary and access adjacent heap metadata, including compressed pointers that reveal address information.
The key advantage is that we separate the speculated branch value (the map pointer) from the leakage-critical data fields.
The map pointer, which must be evicted to trigger misspeculation, resides at the beginning of the object, while the data to be leaked is located several cache lines away.
This separation eliminates the strict split-line alignment requirement. \Cref{fig:leakheap} illustrates the memory layout for our heap-address gadget.

\paragrabf{Pigeonhole Eviction.}
To avoid explicit eviction-set construction, we adapt the pigeonhole-eviction technique proposed by Johnson~\cite{Johnson2021Another}.
In V8, every JavaScript object stores a \textit{map pointer} as its first field, a hidden-class descriptor that JIT-compiled type checks read to verify the object's type.
This map pointer must be evicted to stall branch resolution and open a speculation window.
Rather than constructing an eviction set to flush this value before each gadget invocation, we allocate a large pool of attacker objects and select a random pair per invocation.
With a pool of \SIx{500000} objects, the attacker-object pairs alone span over \SI{40}{\mebi\byte} of V8 heap metadata, far exceeding any LLC.
Any randomly selected pair is therefore guaranteed to be an LLC miss. The \emph{pigeonhole principle} ensures this when the working set exceeds the cache capacity.

The gadget requires an asymmetric cache state: the map pointer must be evicted, while the data fields of the adjacent \texttt{fakeUint32Array}, the target of the transient read, must reside in cache for the transient read to complete before the window closes.
Pigeonhole eviction guarantees the former without any explicit flush or eviction-set traversal.
We prefetch the data fields of the selected pair three times prior to each gadget call to ensure the latter.
Residual incidental map pointer cache hits may occasionally occur, but our nested-loop amplification (\cref{sec:signal_amplification}) amortizes this. A single failed iteration contributes zero signal rather than corrupting the accumulated result.

\paragrabf{Object Pool Allocation.}
We allocate between \SIx{500000} and \SIx{3000000} object triplets, each consisting of two attacker objects and one indirection wrapper.
The two attacker objects together are sized to match the training type (\SIx{188} bytes), ensuring the transient out-of-bounds read lands at the correct offset within the adjacent victim object.
The indirection wrapper holds a reference to the victim typed array, causing V8's garbage collector to promote it adjacent to the speculatively accessed fake typed array in the old-generation heap.
At this scale, triggering garbage collection after each triplet, the conventional promotion approach~\cite{Agarwal2022Spookjs, Schwarzl2022Robust}, would exceed the \SI{30}{\second} initialization limit.
We observe that V8's mark-and-sweep collector traverses the object graph breadth-first.
Referencing alternating attacker and victim objects from a single root array ensures both types are promoted contiguously to the old generation in a single collection pass.
We additionally allocate a warm-up pool of \SIx{2000} objects prior to the main allocation to fill heap fragmentation gaps and improve physical adjacency.
In over \SI{99}{\percent} ($n=10000$) of cases, attacker and victim objects are physically adjacent after promotion.
To avoid non-deterministic RNG latency in the timed section, all random selection indices are pre-computed before the outer measurement loop begins.

\paragrabf{Memory Layout Search.}
After pool allocation, the attacker validates the heap layout before leaking.
The attacker iterates over L1 cache set offsets and measures the PLRU timing signal for a known bit-0 and bit-1 value across \SIx{30} samples per offset.
An offset is accepted if the hit and miss timing distributions are well separated, confirming physical adjacency of attacker and victim objects and a usable PLRU signal.
If no offset yields a clean signal after \SIx{10} attempts, the invocation is aborted.

\begin{figure}[t]
  \centering
  \includegraphics[trim=44 82 136 53, clip, width=\hsize]{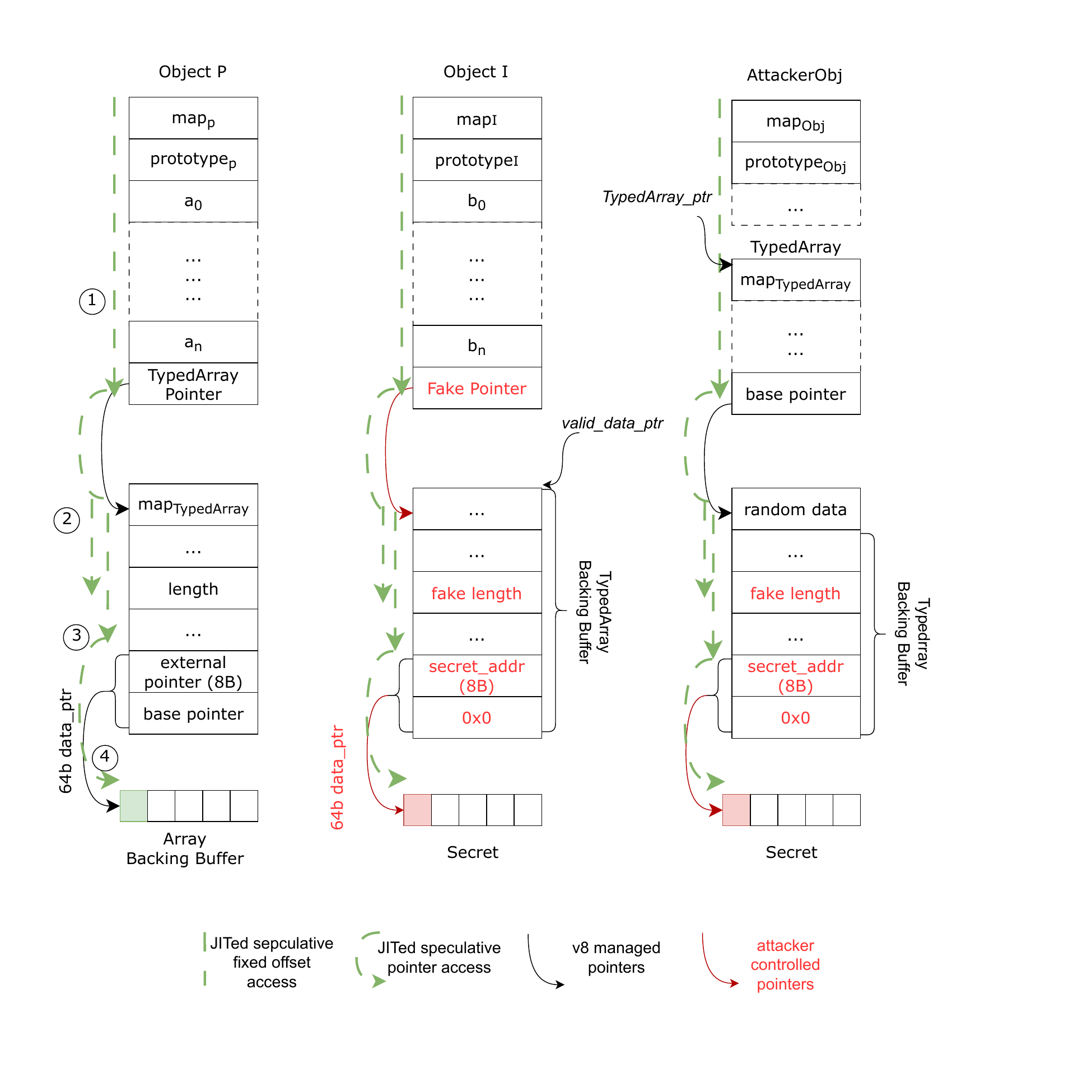}
  \vspace{-0.2cm}
  \caption{Memory layout for the secret-leakage gadget via speculative type confusion. Left: training object~P. Middle: prior approach~\cite{Schwarzl2022Robust} requiring a pre-leaked fake pointer. Right: our approach, writing a crafted fake \texttt{Uint8Array} header into the adjacent backing buffer. Co-locating AttackerObj and TypedArray makes V8 create the equivalent base pointer itself, which removes the fake-pointer setup stage.}
  \label{fig:speculative_type_confusion}
\end{figure}

\paragrabf{Secret-Leakage Gadget.}
The secret-leakage gadget reads data from arbitrary 64-bit addresses within the shared address space, enabling cross-worker data extraction.
Prior secret-leakage gadgets~\cite{Schwarzl2022Robust, Agarwal2022Spookjs} require first leaking a heap address via the heap-address gadget, then fabricating a fake pointer value to drive the transient dereference path (\cref{fig:speculative_type_confusion}, middle).
This dependency chain adds setup latency and increases cumulative error propagation.
We allocate \SIx{500000} pairs of \texttt{AttackerObject} and \texttt{Uint8Array} using the same pigeonhole pool.
Prior to each gadget invocation, we write a fake \texttt{Uint8Array} layout directly into the backing buffer of the selected \texttt{Uint8Array}: a length field at byte offset \SIx{32} and the 64-bit target address as the base pointer at byte offset \SIx{40}.
Under type confusion, the transient path reads the selected \texttt{AttackerObject} as an \texttt{ObjP}.
The \texttt{ptr} field in \texttt{ObjP}'s layout resolves to the adjacent \texttt{Uint8Array}'s backing buffer, where the fake header resides.
The transient path follows the embedded base pointer, the 64-bit target address, and reads \SIx{4} bytes from that location.
Extracting individual bits and encoding them into probe array cache set accesses enables single-bit leakage per invocation.
Note that the \texttt{Uint8Array} backing buffer must not exceed 0x40 bytes, since V8 allocates larger \texttt{Uint8Array} backing buffers in a separate heap region, which breaks the layout assumption.
The attacker applies the same memory layout search as in \cref{sec:spectre_gadget}, retrying up to \SIx{5}$\times$.

\Cref{fig:speculative_type_confusion} illustrates the memory layout, and \cref{lst:origGadget} sketches the corresponding training and leakage code.
The branch in \texttt{spectreGadget} is mistrained on \texttt{ObjP} instances so that, when invoked on the adjacent \texttt{Uint8Array}-backed object, the JIT speculatively follows the \texttt{obj.ptr[0]} dereference and emits the encoded cache access consumed by our PLRU amplifier (\cref{sec:signal_amplification}).
On the AMD EPYC CPUs deployed in \CFWorkers, training only the local branch is insufficient: TAGE-style predictors~\cite{Seznec2016} index on long global history, so we additionally invoke \texttt{wasm\_plru} during training to match the global branch context observed during leakage (\SIx{20} local-history and \SIx{5} global-history training rounds per leakage call).
Compared to prior variants that require constructing a fake TypedArray at a pre-leaked heap address~\cite{Agarwal2022Spookjs}, our design removes a dependency stage entirely: the target address is injected directly into the adjacent \texttt{Uint8Array} backing buffer, requiring no prior address leak to bootstrap.
The secret-leakage gadget can be replaced with any other 64-bit Spectre gadget~\cite{Kocher2019,Canella2019A,Mcilroy2019Spectre,Kim2024Tiktag,Wikner2022Spring,Koruyeh2018spectrersb,Maisuradze2018spectrersb,Agarwal2022Spookjs} that produces a transient read of an attacker-controlled address.

\paragrabf{Speculative-Window Enlargement.}
A successful exfiltration requires that both the dereference of \texttt{obj.ptr} and the subsequent load from the embedded 64-bit address complete before the misprediction is resolved.
We extend the transient window in two ways.
First, the \texttt{instanceof} check in \texttt{spectreGadget} is preceded by a sequence of long-latency arithmetic operations on a value derived from the map pointer, so that branch resolution stalls until that chain retires.
Second, prior to entering the gadget we issue several dependent integer divisions whose result feeds the branch condition.
These divisions occupy the AMD scheduler's execution ports for tens of cycles.
They widen the speculation window enough for the two-level pointer chase to retire its cache effects.
We empirically tune the number of arithmetic operations per call, typically 4 to 8, until the encoded cache access becomes observable through the PLRU amplifier.
Lower counts collapse the window and the leakage drops to noise.
Excessive counts only reduce throughput without improving accuracy.

\paragrabf{Encoding and Decoding.}
For each leaked bit, the gadget masks out a single bit from the transiently read 4-byte value and uses it to select between two disjoint cache sets in the probe array.
The PLRU amplifier (\cref{sec:signal_amplification}) then accumulates the timing asymmetry between these two sets into a remote-timer-observable difference.
The attacker reconstructs a full 64-bit address bitwise across \SIx{32} invocations of the heap-address gadget followed by \SIx{32} invocations of the secret-leakage gadget.
Each per-bit decision uses the threshold calibrated as described in \cref{sec:signal_amplification}.

\begin{listing}[t]
  \begin{lstlisting}[caption={Speculative type confusion training and leaking.},label={lst:origGadget},language=JavaScript,style=customjs]
spectreGadget() { 
if (obj instanceof ObjP) { leak(obj.ptr[0]); } 
}
wasm_plru() {
  // step 1: setup cache for PLRU
  // step 2: call spectreGadget
  // step 3: PLRU amplification
}
// Training: mistrain branch predictor
for (let j = 0; j < 20; j++) {
  spectreGadget(); // local history
}
for (let j = 0; j < 5; j++) {
  wasm_plru(); // global history
}
// Leaking
wasm_plru();\end{lstlisting}
  \vspace{-0.6cm}
\end{listing}

\begin{tcolorbox}[boxsep=1pt,left=2pt,right=2pt,top=1pt,bottom=1pt]
  \textbf{C3.2: Robustness.}
  Pigeonhole eviction replaces explicit eviction-set construction with random selection from a large object pool, eliminating the per-invocation setup cost and its \SI{50}{\percent} failure rate.
  The heap-address gadget removes the cacheline-alignment requirement via speculative type confusion, and our secret-leakage gadget removes the fake-pointer dependency.
\end{tcolorbox}

\section{Evaluation}\label{sec:evaluation}
In this section, we demonstrate the practicality of our presented attack primitives in the production environment of \CFWorkers. Our attack significantly outperforms the reported maximum leakage rate of \SI{120}{\bit/\hour}~\cite{Schwarzl2022Robust} with a leakage rate of up to \SI{12}{\bit/\second} (\SI{43200}{\bit/\hour}).

\paragrabf{Setup.}
We run all our experiments with the (\textbf{WS+Sbx}) timer, co-located on the same machine.
The target systems were running on Linux servers equipped with AMD EPYC CPUs (Zen2 and Zen3).
\CFWorkers operates a highly homogeneous server fleet, with successive generations uniformly deploying AMD EPYC processors~\cite{CFGen11,CFGen12,CFGen13}.
This hardware uniformity benefits the attacker, as microarchitectural assumptions such as cache-replacement policies and scheduler behavior remain consistent across machines.
Consequently, an attack gadget calibrated on one machine is likely to behave identically on another, removing the need for per-machine parameter tuning.
Note that the following experiments running on the production system are intentionally targeting memory regions owned by a crafted victim isolate to not accidentally leak customers data.
To get comparable results, we pin our measurements for both attacker and victim scripts to a specific machine.
We run the experiments at night on the machine, with a CPU utilization between \SIx{10} and \SIx{25} percent to observe the best possible results.

\begin{figure}[t]
    \centering
    \begin{subfigure}{\hsize}%
      \begin{tikzpicture}
  \begin{axis}[
    style={font=\footnotesize},
    xlabel={Time (HH:MM)},
    ylabel={Rate},
    y label style={align=center,text width=2cm},
    width=\hsize,
    height=3cm,
    xmin=13,
    xmax=80,
    ymax=1,
          xtick={24,36,48,60,72
        },
              xticklabels={
           00:00, 12:00, 00:00, 12:00, 00:00
        },
    legend style={at={(.8,1.3)}, anchor=north east, legend columns=2, font=\footnotesize,draw=none,fill=none}
    ]

    \addplot[red,smooth,thick] table [x index=0,y index=1,col sep=comma] {data/warsaw_stockholm_win10_success.csv};
    \addplot[blue,dashed] table [x index=0,y index=1,col sep=comma] {data/warsaw_stockholm_win10_utilization.csv};
    \legend{Success Rate, CPU Utilization}
  \end{axis}

\end{tikzpicture}%
    \end{subfigure}
    \vspace{-0.35cm}
    \caption{Success rate of initial memory layout step on a production system using a remote timer. We consider the stage to be successful if a run starts leaking memory.}
    \label{fig:success_rate_memory_layout}
\end{figure}
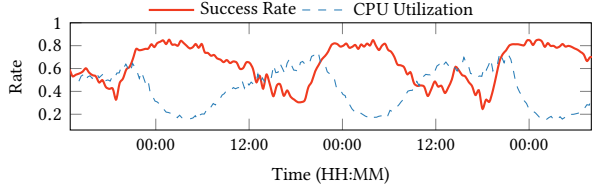

\paragrabf{Memory Layout.}
To indicate the strong variance on the production system, we performed a first experiment on the success rate of identifying a stable memory layout.
We measure how often a stable memory layout ready for one of our gadgets succeeds. 
We run this experiment for a period of roughly three days. \Cref{fig:success_rate_memory_layout} illustrates the success rate of the experiment and the dashed line illustrates the CPU utilization on that specific machine during that time period.
The stage is considered successful if it successfully enters the next stage of leaking the compressed pointer.
There is a direct correlation between the success rate of our experiment and the systems CPU utilization.
The system utilization thus influences the success rate of each of our presented experiments.
An attacker can maximize leakage rate by targeting off-peak hours, though slower attacks remain feasible at any time of day even under high load.

\paragrabf{Timer Stability.}
We collect \SIx{253} ground-truth sessions over \SIx{28} hours on the production system.
The per-session miss$-$hit difference stays strictly positive, with a median of \SI{1.17}{\milli\second} ($n = 253, \sigma_{\mu} = 4.75\%$).
We conclude that hit/miss distinguishability is preserved across time.
The absolute throughput follows a load-dependent diurnal pattern, \cf \Cref{app:timer_stability}.

\paragrabf{Classifying a bit.}
We classify leaked bits in remote Spectre attacks by first building an empirical baseline from 100 architecturally reachable, \emph{known} values to capture the distributions of measured ``0'' and ``1'' responses.
From that baseline, we determine a classification threshold as the 35th percentile of the one-bit timing distribution, which empirically produced the most robust separation.
For each leaked bit we then perform a two-sided hypothesis test using that threshold.
We count how many measured samples fall above the threshold when testing for a one and how many fall above it when testing for a zero, and the majority determines the inferred bit.
Aggregating these per-bit decisions across a byte yields the byte classification.

\paragrabf{Bypassing \mitigation with Durable Objects.}
\mitigation enforces process isolation \textbf{after} a script execution finished. 
This means, that as long as the Durable Object is kept alive it can leak without being interrupted. 
Due to the long-living nature of Durable Objects, not process isolating is a major implementation flaw when trying to detect on-going attacks.
Moreover, \mitigation relies on normalizing the amount of branch accesses by the amount of iTLB accesses.
As our WebSocket-based remote timer is frequently called, the amount of code being used between a sample is increased, thereby lowering the ratio between an event normalized by iTLB accesses.

\paragrabf{Leaking the isolate root.}
V8 uses pointer compression to reduce memory consumption and improve cache locality.
In V8, all heap objects reside in a \SI{4}{\gibi\byte} region called the isolate heap, allowing 64-bit pointers to be represented as 32-bit signed offsets from a base pointer, the upper half of virtual address, referred to as the isolate root~\cite{v8pointercompression}.
This technique is used extensively for object properties, internal fields, and array element pointers.

We design a \worker that uses the speculative type confusion gadget described in \cref{sec:spectre_gadget} to leak the isolate root.
The speculative type confusion gadget leverages a type confusion between a \texttt{Uint8Array} into a \texttt{Uint32Array}.
The gadget reads 4 bytes with one memory access at the given index and masks out the bit to be leaked. 
To verify if our leaked isolate root is correct, we place known random values next to our target pointer, which we leak in addition.
In addition, leveraging an internal binding, we can verify whether our leaked address is correct.
We repeat the attack \SIx{1000} times. 
A single run is considered successful if we correctly leak the isolate root. 
We configure the attack to run with a window size of \SI{5}, \SI{8500} inner loops and \SI{25} outer loops, achieving a leakage rate of \SI{12}{\bit/\second} ($n = 100, \sigma_{\mu} = 16.75\%$) with an average accuracy of \SI{99.60}{\percent}.

\paragrabf{Memory Layout.}
We leak the full isolate address over \SIx{10} consecutive redeployments.
We observe a linear allocation of isolate roots with a \SI{2}{\gibi\byte} offset.
Moreover, the bump allocator in V8 keeps object offsets stable across executions of the same script.
We conclude that both properties ease the search for valuable addresses, \cf \Cref{app:memory_layout}.

\paragrabf{Leaking from the vDSO.}
To test leaking from a 64-bit address, \CFWorkers team added an experimental internal debug binding for our experiments that returns the address of the vDSO section.
We design a \worker that uses the gadget to leak arbitrary data.
Our attack leaks \SI{128}{\byte} from the vDSO. 
We start at offset \texttt{0xe20}, where the string \textit{gettimeofday} is located.

\begin{figure}[t]
    \centering
    \begin{subfigure}{\hsize}%
      \includegraphics[width=\hsize]{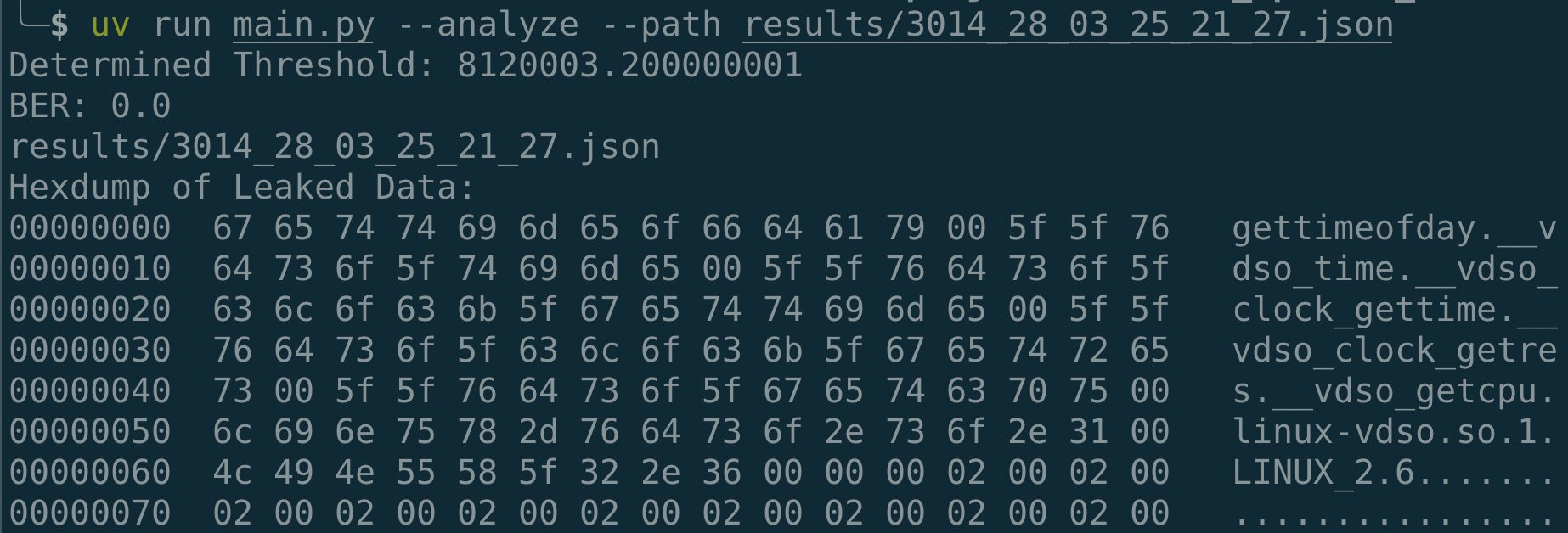}%
    \end{subfigure}
    \vspace{-0.35cm}
    \caption{Leakage of vDSO data from production system with no bit errors.}
    \label{fig:vdso}
\end{figure}

Our parameterization on the production system is a window size of \SIx{10}, \SIx{8500} inner loops and \SIx{25} outer loops.
With this configuration and the \textbf{WS+Sbx} timer, we achieve a leakage rate of up to \SI{12}{\bit/\second} ($n = 100, \sigma_{\mu} = 30.14\%$) with an average accuracy of \SI{99.16}{\percent}. 
\Cref{fig:vdso} displays the leaked bytes of our attack with perfect results. 

\paragrabf{Leaking from another Worker.}
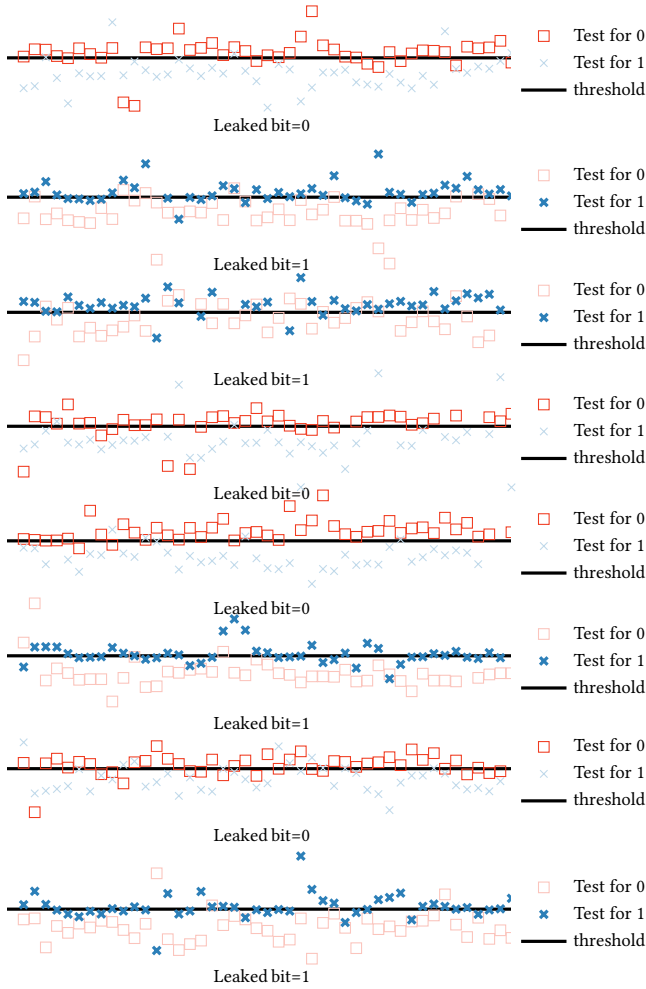
\begin{figure}[t]
    \centering
    \begin{subfigure}{\hsize}%
      \begin{tikzpicture}
  \begin{axis}[
    style={font=\footnotesize},
    xlabel={Leaked\,bit=0},
    y label style={align=center,text width=2cm},
    width=\hsize,
    height=3cm,
    ymin=4500000,
    ymax=13000000,
    xmin=-2,
    xmax=45,
    axis line style={draw=none},
    tick style={draw=none},
    xtick=\empty,
    ytick=\empty,
    legend style={at={(1.12,0)}, anchor=south, font=\footnotesize, draw=none}
  ]

    \addplot[red!100,mark=square,only marks] table[x expr=\coordindex,y index=0] {data/leak_byte_e/1_7_0_8585398.txt};
    \addlegendentry{Test for 0}

    \addplot[blue!30,mark=x,only marks] table[x expr=\coordindex,y index=0] {data/leak_byte_e/1_7_1_8585398.txt};
    \addlegendentry{Test for 1}

    \addplot[black,very thick] coordinates {(-1.5,8585398) (45.5,8585398)};
    \addlegendentry{threshold}

  \end{axis}
\end{tikzpicture}%
    \end{subfigure}
    \vspace{-0.35cm}
    \begin{subfigure}{\hsize}%
      \begin{tikzpicture}
  \begin{axis}[
    style={font=\footnotesize},
    xlabel={Leaked\,bit=1},
    y label style={align=center,text width=2cm},
    width=\hsize,
    height=3cm,
    ymin=4500000,
    ymax=13000000,
    xmin=-2,
    xmax=45,
    axis line style={draw=none},
    tick style={draw=none},
    xtick=\empty,
    ytick=\empty,
    legend style={at={(1.12,0)}, anchor=south, font=\footnotesize, draw=none}
  ]

    \addplot[red!30,mark=square,only marks] table[x expr=\coordindex,y index=0] {data/leak_byte_e/1_6_0_8585398.txt};
    \addlegendentry{Test for 0}

    \addplot[blue!100,very thick,mark=x,only marks] table[x expr=\coordindex,y index=0] {data/leak_byte_e/1_6_1_8585398.txt};
    \addlegendentry{Test for 1}

    \addplot[black,very thick] coordinates {(-1.5,8585398) (45.5,8585398)};
    \addlegendentry{threshold}

  \end{axis}
\end{tikzpicture}%
    \end{subfigure}
    \vspace{-0.35cm}
    \begin{subfigure}{\hsize}%
      \begin{tikzpicture}
  \begin{axis}[
    style={font=\footnotesize},
    xlabel={Leaked\,bit=1},
    y label style={align=center,text width=2cm},
    width=\hsize,
    height=3cm,
    ymin=4500000,
    ymax=13000000,
    xmin=-2,
    xmax=45,
    axis line style={draw=none},
    tick style={draw=none},
    xtick=\empty,
    ytick=\empty,
    legend style={at={(1.12,0)}, anchor=south, font=\footnotesize, draw=none}
  ]

    \addplot[red!30,mark=square,only marks] table[x expr=\coordindex,y index=0] {data/leak_byte_e/1_5_0_8585398.txt};
    \addlegendentry{Test for 0}

    \addplot[blue!100,very thick,mark=x,only marks] table[x expr=\coordindex,y index=0] {data/leak_byte_e/1_5_1_8585398.txt};
    \addlegendentry{Test for 1}

    \addplot[black,very thick] coordinates {(-1.5,8585398) (45.5,8585398)};
    \addlegendentry{threshold}

  \end{axis}
\end{tikzpicture}%
    \end{subfigure}
    \vspace{-0.35cm}
    \begin{subfigure}{\hsize}%
      \begin{tikzpicture}
  \begin{axis}[
    style={font=\footnotesize},
    xlabel={Leaked\,bit=0},
    y label style={align=center,text width=2cm},
    width=\hsize,
    height=3cm,
    ymin=4500000,
    ymax=13000000,
    xmin=-2,
    xmax=45,
    axis line style={draw=none},
    tick style={draw=none},
    xtick=\empty,
    ytick=\empty,
    legend style={at={(1.12,0)}, anchor=south, font=\footnotesize, draw=none}
  ]

    \addplot[red!100,mark=square,only marks] table[x expr=\coordindex,y index=0] {data/leak_byte_e/1_4_0_8585398.txt};
    \addlegendentry{Test for 0}

    \addplot[blue!30,mark=x,only marks] table[x expr=\coordindex,y index=0] {data/leak_byte_e/1_4_1_8585398.txt};
    \addlegendentry{Test for 1}

    \addplot[black,very thick] coordinates {(-1.5,8585398) (45.5,8585398)};
    \addlegendentry{threshold}

  \end{axis}
\end{tikzpicture}%
    \end{subfigure}
    \vspace{-0.35cm}
    \begin{subfigure}{\hsize}%
      \begin{tikzpicture}
  \begin{axis}[
    style={font=\footnotesize},
    xlabel={Leaked\,bit=0},
    y label style={align=center,text width=2cm},
    width=\hsize,
    height=3cm,
    ymin=4500000,
    ymax=13000000,
    xmin=-2,
    xmax=45,
    axis line style={draw=none},
    tick style={draw=none},
    xtick=\empty,
    ytick=\empty,
    legend style={at={(1.12,0)}, anchor=south, font=\footnotesize, draw=none}
  ]

    \addplot[red!100,mark=square,only marks] table[x expr=\coordindex,y index=0] {data/leak_byte_e/1_3_0_8585398.txt};
    \addlegendentry{Test for 0}

    \addplot[blue!30,mark=x,only marks] table[x expr=\coordindex,y index=0] {data/leak_byte_e/1_3_1_8585398.txt};
    \addlegendentry{Test for 1}

    \addplot[black,very thick] coordinates {(-1.5,8585398) (45.5,8585398)};
    \addlegendentry{threshold}

  \end{axis}
\end{tikzpicture}%
    \end{subfigure}
    \vspace{-0.35cm}
    \begin{subfigure}{\hsize}%
      \begin{tikzpicture}
  \begin{axis}[
    style={font=\footnotesize},
    xlabel={Leaked\,bit=1},
    y label style={align=center,text width=2cm},
    width=\hsize,
    height=3cm,
    ymin=4500000,
    ymax=13000000,
    xmin=-2,
    xmax=45,
    axis line style={draw=none},
    tick style={draw=none},
    xtick=\empty,
    ytick=\empty,
    legend style={at={(1.12,0)}, anchor=south, font=\footnotesize, draw=none}
  ]

    \addplot[red!30,mark=square,only marks] table[x expr=\coordindex,y index=0] {data/leak_byte_e/1_2_0_8585398.txt};
    \addlegendentry{Test for 0}

    \addplot[blue!100,very thick,mark=x,only marks] table[x expr=\coordindex,y index=0] {data/leak_byte_e/1_2_1_8585398.txt};
    \addlegendentry{Test for 1}

    \addplot[black,very thick] coordinates {(-1.5,8585398) (45.5,8585398)};
    \addlegendentry{threshold}

  \end{axis}
\end{tikzpicture}%
    \end{subfigure}
    \begin{subfigure}{\hsize}%
      \begin{tikzpicture}
  \begin{axis}[
    style={font=\footnotesize},
    xlabel={Leaked\,bit=0},
    y label style={align=center,text width=2cm},
    width=\hsize,
    height=3cm,
    ymin=4500000,
    ymax=13000000,
    xmin=-2,
    xmax=45,
    axis line style={draw=none},
    tick style={draw=none},
    xtick=\empty,
    ytick=\empty,
    legend style={at={(1.12,0)}, anchor=south, font=\footnotesize, draw=none}
  ]

    \addplot[red!100,mark=square,only marks] table[x expr=\coordindex,y index=0] {data/leak_byte_e/1_1_0_8585398.txt};
    \addlegendentry{Test for 0}

    \addplot[blue!30,mark=x,only marks] table[x expr=\coordindex,y index=0] {data/leak_byte_e/1_1_1_8585398.txt};
    \addlegendentry{Test for 1}

    \addplot[black,very thick] coordinates {(-1.5,8585398) (45.5,8585398)};
    \addlegendentry{threshold}

  \end{axis}
\end{tikzpicture}%
    \end{subfigure}
    \vspace{-0.35cm}
    \begin{subfigure}{\hsize}%
      \begin{tikzpicture}
  \begin{axis}[
    style={font=\footnotesize},
    xlabel={Leaked\,bit=1},
    y label style={align=center,text width=2cm},
    width=\hsize,
    height=3cm,
    ymin=4500000,
    ymax=13000000,
    xmin=-2,
    xmax=45,
    axis line style={draw=none},
    tick style={draw=none},
    xtick=\empty,
    ytick=\empty,
    legend style={at={(1.12,0)}, anchor=south, font=\footnotesize, draw=none}
  ]

    \addplot[red!30,mark=square,only marks] table[x expr=\coordindex,y index=0] {data/leak_byte_e/1_0_0_8585398.txt};
    \addlegendentry{Test for 0}

    \addplot[blue!100,very thick,mark=x,only marks] table[x expr=\coordindex,y index=0] {data/leak_byte_e/1_0_1_8585398.txt};
    \addlegendentry{Test for 1}

    \addplot[black,very thick] coordinates {(-1.5,8585398) (45.5,8585398)};
    \addlegendentry{threshold}

  \end{axis}
\end{tikzpicture}%
    \end{subfigure}
    \vspace{1pt}
    \caption{Bitwise classification results for the first byte (the letter `e', \texttt{0b01100101}) of the leaked JWT token. Each subplot shows the separation between `0' and `1' tests for a single bit.}
    \label{fig:byte_leakage}
\end{figure}
We demonstrate the full end-to-end attack by leaking a JWT token placed in a co-located victim \worker.
The attack chain proceeds as follows.
First, the attacker uses the heap-address gadget to leak its own isolate root.
Next, the attacker uses the secret-leakage gadget to follow internal pointer chains relative to that root.
This enables traversal into the shared process address space and eventual access to the victim \worker's heap.
We target a JWT token stored in the victim's memory and leak it byte by byte.
\Cref{fig:byte_leakage} illustrates the bitwise classification for the first byte of the leaked token.
The measured bit-classification accuracy and leakage rate are in line with the vDSO experiment, as the same gadgets and amplification parameters are used.

\section{Mitigations}\label{sec:mitigations}
\mitigation~\cite{Schwarzl2022Robust} monitors performance counters per \worker script, normalizes them by iTLB accesses, and process-isolates scripts that exceed a detection threshold.
The deployment in \CFWorkers reveals two implementation flaws that our attack exploits.
First, isolation is enforced only after a script execution completes.
Since Durable Objects never complete their invocation while keep-alive signals are received, the attacker leaks data throughout a single uninterrupted execution and is never isolated.
Second, the normalization metric is not robust to I/O-heavy code paths.
WebSocket timer calls introduce substantial iTLB activity independent of the Spectre gadget, suppressing the normalized ratio below the detection threshold.
We conclude that both flaws are fundamental limitations of the detection approach rather than implementation oversights.
An attacker with control over the I/O pattern can always inflate iTLB activity to evade normalization-based detection.
A robust detection requires real-time monitoring during execution rather than post-hoc isolation, and a metric invariant to the monitored script's I/O workload.
The detection signals could be extended by focusing on the exfiltration stage \ie timing information.
\paragrabf{V8 Sandbox.}
The basic idea of the new V8 sandbox is to eliminate remaining raw (64-bit) pointers~\cite{v8sandbox}. 
This would reduce the attack surface, in case an attacker has an arbitrary read and write primitive within the sandbox, by restricting it to only the heap.  
Thus, an attacker cannot corrupt memory outside of the pointer “cage”.
Heap pointers become 32-bit offsets into a \SI{4}{\gibi\byte} pointer cage.
Buffer pointers, such as TypedArray backing stores, become 33-bit offsets that address the full \SI{8}{\gibi\byte} sandbox~\cite{CloudflareSandboxHardening2025}.
\CFWorkers has deployed the sandbox by introducing per-isolate sandbox groups into V8~\cite{CloudflareSandboxHardening2025}.
We can confirm that the presented gadget is mitigated if the V8 sandbox is enabled by testing locally.
The type confusion used in the attack is not possible anymore, as there is no raw 64-bit pointer used.
However, there are many other objects that are not yet ported to the V8 sandbox, and could be leveraged for crafting gadgets.  
Moreover, other Spectre variants might still be exploitable, given the vast amount of potential for crafting Spectre gadgets.
We conclude that the V8 sandbox hardens the crafting of simple 64-bit speculative type confusions, however, it does not remove the potential of Spectre gadgets.
\paragrabf{In-Process Isolation.}
Google previously gave up on in-process Spectre defense, and implemented full process isolation for each origin in Chrome~\cite{Reis2019Siteisolation,Mcilroy2019Spectre}.
Still, in-process isolation at scale, could play a crucial part in protecting the tenants from in-process Spectre.
Kiriansky~\cite{Kiriansky2018dawg} demonstrated that hardware-assisted memory protection mechanisms like Intel MPK can mitigate Spectre attacks. 
Various MPK-based schemes were proposed to provide efficient in-process isolation~\cite{Schrammel2020Donky,Blair2023ThreadLock,Hedayati2019Hodor,Schrammel2022Jenny}.
The major drawback of MPK is the limited number of protection domains, with up to \SIx{15} keys available per process. 
VDom~\cite{Yuan2023Vdom} overcomes the security domain implementation by leveraging address-space identifiers (ASIDs).
Unterguggenberger~\etal\cite{Unterguggenberger2024TMEBox} presented an approach leveraging Intel-TME for in process isolation supporting up to 32K tenants per process.
While these approaches look promising the overhead for a strongly multithreaded environment such as \CFWorkers is unclear.
\CFWorkers deployed MPK-based in-process isolation as a coordinated mitigation~\cite{CloudflareSandboxHardening2025}.
Modern x64 CPUs expose up to \SIx{15} protection keys.
V8 reserves a few keys for its JIT compilers, leaving approximately \SIx{12} keys for isolation.
\CFWorkers assigns a protection key to each isolate heap.
A cross-isolate access traps in hardware unless the accessing thread holds the matching key.
For our attack, the transient read of the secret-leakage gadget now targets a differently-keyed heap region and faults unless the attacker isolate shares the victim key.
With random key assignment, this traps approximately \SI{92}{\percent} of cross-isolate accesses, corresponding to \SIx{11} out of \SIx{12} keys.
To reach complete coverage, \CFWorkers combines the keys with the V8 sandbox and places each sandbox under a strictly rotating key.
No sandbox within a \SI{32}{\gibi\byte} window shares a key, so neighboring sandboxes act as each other's guard regions.
The residual risk is an attacker who controls the PKRU register, requiring arbitrary code execution~\cite{Connor2020Pku}.
Our attack demonstrates the necessity of hardware-enforced isolation such as MPK.
The software-only countermeasures of \CFWorkers, restricted timers and performance-counter detection, prove insufficient against state-of-the-art amplification and evasion techniques.
We conclude that hardware-assisted protection is the most effective mitigation against in-process Spectre attacks in a shared-process environment.

\paragrabf{Secrets Proxy.}
LeakLess~\cite{Rostamipoor2025NDSS} introduces an innovative approach to protecting secrets in Function-as-a-Service (FaaS) environments. 
The primary goal of LeakLess is to store secrets outside the language-level isolated process, instead placing them in a separate I/O module. 
This module can either run in a distinct process or on an entirely different machine.
To enable this, secrets must be annotated.
The module then decrypts and inserts them into requests as needed, for instance when a subrequest requires a bearer token.
While this approach offers enhanced protection for secrets in \CFWorkers, it requires manual annotation by developers to flag secrets.
Moreover, the scalability of this approach for \CFWorkers is unclear.
The entire request object may contain sensitive information, and certain operations may process these secrets directly within the worker.

\section{Conclusion}\label{sec:conclusion}
In this paper, we reconsidered the security of \CFWorkers with respect to remote Spectre attacks.
We explored the attackers' possibilities to co-locate remote timers in \CFWorkers.
Our timers achieve a median accuracy in the tens of microseconds ($\sigma$ \SI{0.5}{\milli\second} to a few milliseconds) under production load.
There is no way of mitigating a remote timer if an attacker is allowed to connect to remote hosts.
In combination with two speculative type confusion gadgets, we mounted a remote Spectre attack in the production system of \CFWorkers, leaking a JWT token from a co-located victim worker at up to \SI{12}{\bit/\second} ($n = 100, \sigma_{\mu} = 30.14\%$) with an accuracy of \SI{99.16}{\percent}, outperforming the previous attack by a factor of 360$\times$.
The \CFWorkers team found no indicators of the vulnerability being actively exploited.
\CFWorkers mitigated the attack in a coordinated effort by integrating the V8 Sandbox limiting transient access to 64-bit pointers, improving the detection capabilities of \mitigation, and deploying MPK-based in-process isolation.
Our attack demonstrates the necessity of hardware-enforced isolation, as the software-only countermeasures of \CFWorkers proved insufficient against state-of-the-art amplification and evasion techniques.

\newpage
\bibliographystyle{ACM-Reference-Format}
\bibliography{references}

\appendix
\crefalias{section}{appendix}

\section{Nested-Loop Amplification Scaling}
\label{app:amplification}

The nested-loop gadget of \cref{lst:fineGrainTimer} exposes two repetition counts.
The inner loop (\texttt{INNER\_REP}) drives the PLRU oscillation within a single iteration.
The outer loop (\texttt{OUTER\_REP\_NUM}) repeats setup, transient leak, and amplification.

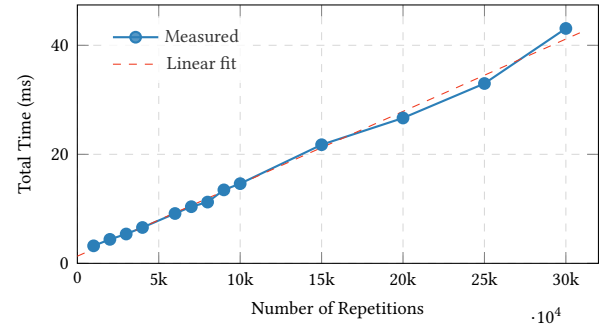
\begin{figure}[H]
  \centering
\begin{tikzpicture}
  \begin{axis}[
    style={font=\footnotesize},
    xlabel={Number of Repetitions},
    ylabel={Total Time (ms)},
    width=\hsize,
    height=5cm,
    xmin=0,
    xmax=32000,
    ymin=0,
    grid=major,
    grid style={dashed,gray!30},
    xtick={0,5000,10000,15000,20000,25000,30000},
    xticklabels={0,5k,10k,15k,20k,25k,30k},
    legend style={at={(0.05,0.95)}, anchor=north west, font=\footnotesize, draw=none, fill=white, fill opacity=0.8},
  ]

    \addplot[blue, thick, mark=*, mark size=2pt]
      table [x=inner_loops, y=total_ms, col sep=comma]
      {data/amplification_scaling.csv};

    \addplot[red, dashed, thin, domain=0:31000, samples=2] {0.001328*x + 1.3102};

    \legend{Measured, Linear fit}

  \end{axis}
\end{tikzpicture}%
  \vspace{-0.35cm}
  \caption{Total timing as a function of inner-loop repetitions for the nested-loop PLRU amplification.}
  \label{fig:amplification_scaling}
\end{figure}

\Cref{fig:amplification_scaling} shows the total measured timing as a function of the inner-loop repetitions.
We measure a linear relation over the entire evaluated range.
We conclude that the amplification factor is a directly controllable parameter.
An attacker chooses \texttt{INNER\_REP} to raise the timing difference above the resolution of the available remote timer.
The attacker pays for this choice only in wall-clock time.

\section{Long-Term Timer Stability}
\label{app:timer_stability}

To evaluate whether the timing signal remains distinguishable over an extended period, we measure the median cache-hit and cache-miss timing on the production system.
We collect \SIx{253} ground-truth measurement sessions over \SIx{28} hours.
Each session records rdtscp cycle counts for cache hits (\texttt{lastTest0\_1}, \texttt{lastTest1\_0}) and cache misses (\texttt{lastTest0\_0}, \texttt{lastTest1\_1}).

\begin{figure}[H]
    \centering
    \begin{subfigure}{\hsize}%
      \begin{tikzpicture}
    \begin{axis}[
        style={font=\footnotesize},
        xlabel={Time (HH:MM)},
        ylabel={Miss $-$ Hit [ms]},
        width=\hsize,
        height=3cm,
        ymin=0, ymax=4.5,
        xmin=8, xmax=36,
        xtick={12,24,36},
        xticklabels={12:00, 00:00, 12:00},
        axis x line=bottom,
        axis y line=left,
        axis background/.style={fill=none},
        legend style={at={(0.5,1.03)}, anchor=south, draw=none, fill=none, font=\footnotesize},
        legend columns=-1,
        legend cell align=left,
    ]
        \draw[gray, dashed, line width=0.6pt] ({axis cs:24,0}) -- ({axis cs:24,4.45});
        \node[gray, font=\tiny, anchor=north east] at (axis cs:24.1,4.45) {midnight};
        \addplot[black, solid, thick, each nth point=1]
            table [x index=0, y index=1, col sep=comma] {data/timeseries_gap.csv};
        \addlegendentry{Median timing difference}
    \end{axis}
\end{tikzpicture}%
    \end{subfigure}
    \vspace{-0.35cm}
    \caption{Per-session miss$-$hit timing difference over \SIx{28} hours on the production system. The difference stays strictly positive across the entire window.}
    \label{fig:timeseries_remote_timer}
\end{figure}
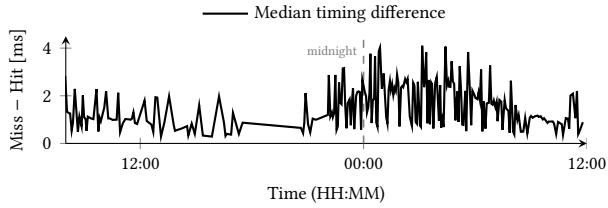

\Cref{fig:timeseries_remote_timer} plots the per-session miss$-$hit timing difference over the full observation window.
The difference remains strictly positive across the entire window.
We measure a median of \SI{1.17}{\milli\second} ($n = 253, \sigma_{\mu} = 4.75\%$).

The measurements also show a diurnal pattern.
During afternoon hours, the production system is under heavy tenant load.
Fewer sessions complete successfully, and the per-session difference shrinks toward the noise floor.
After midnight, contention drops and the difference grows correspondingly larger.
We conclude that the absolute leakage throughput is load-dependent.
Still, the underlying signal never disappears.

\section{Isolate and Object Memory Layout}
\label{app:memory_layout}

We use the attack to verify the address allocation of isolate roots.
We redeploy the script after each execution to create a new isolate.
\Cref{fig:linear_allocation} shows the isolate roots we leak over \SIx{10} consecutive runs.
We observe a linear allocation with a \SI{2}{\gibi\byte} offset between the newly allocated isolates.

Moreover, we analyze the addresses of local and global objects within an isolate root.
The addresses stay the same across script executions until we redeploy the script.
The bump allocator in V8 places offsets within \SI{256}{\kibi\byte} pages at the same positions.

\begin{figure}[H]
  \centering
  \begin{subfigure}{\hsize}%
    \resizebox{\hsize}{!}{
\begin{tikzpicture}[every node/.style={font=\ttfamily}]
  \foreach[count=\i from 0] \addr in {0x10ea,0x10ec,0x10ee,0x10f0,0x10f2,0x10f4,0x10f6,0x10f8,0x10fa,0x10fc} {
    \pgfmathtruncatemacro{\row}{div(\i,5)}
    \pgfmathtruncatemacro{\col}{mod(\i,5)}
    \node[draw, minimum width=1.5cm, minimum height=0.7cm]
      at (1.55*\col, -0.8*\row) {\addr};
  }
\end{tikzpicture}
}%
  \end{subfigure}
  \vspace{-0.35cm}
  \caption{Leaked isolate roots, indicating linear isolate allocation within \SI{2}{\gibi\byte}. Leaked objects sit at a fixed offset from the isolate root due to bump allocation, e.g., 0x10ea00077e24.}
  \label{fig:linear_allocation}
\end{figure}

\end{document}